\documentclass[aps,pra,twocolumn,superscriptaddress]{revtex4-2}
\usepackage{amsmath,amssymb}
\usepackage{graphicx}
\usepackage{booktabs}
\usepackage{tikz}
\usetikzlibrary{arrows.meta,decorations.pathmorphing,positioning,calc}
\usepackage[colorlinks=true,linkcolor=blue,citecolor=blue]{hyperref}

\makeatletter
\def\selectlanguage#1{}
\makeatother

\begin{document}
\title{Oscillation-Free Unconventional Photon Blockade from Quadrature Driving of a Kerr Dimer}

\author{H.~Ohadi}
\email{ho35@st-andrews.ac.uk}
\affiliation{School of Physics and Astronomy,
  University of St Andrews, St Andrews KY16 9SS, United Kingdom}

\begin{abstract}
We demonstrate unconventional photon blockade in a symmetric Kerr dimer
driven with equal-amplitude fields at a $90^\circ$ phase difference.
The minimum inter-cavity coupling is $J_{\min} = \gamma/4$ at a Kerr nonlinearity
$U \ll \gamma$ achievable in standard photonic molecules.
The quadrature-driven site emits strongly antibunched light with a smooth,
oscillation-free second-order correlator directly resolvable with standard
detectors.  The scheme operates under continuous-wave and pulsed excitation, and
fabrication disorder can be fully compensated by re-tuning the drive phase and amplitude ratio,
removing the need for post-fabrication cavity trimming.
\end{abstract}

\maketitle

\begin{figure}[t]
\centering
\begin{tikzpicture}[
  cavity/.style={rectangle, rounded corners=5pt, draw=black!70, thick,
                 fill=blue!8, minimum width=1.cm, minimum height=1.cm},
  photon/.style={-{Stealth[length=5pt]}, decorate,
                 decoration={snake, amplitude=2pt, segment length=5pt,
                             post length=4pt}, thick},
  arr/.style={-{Stealth[length=5pt]}, thick},
  font=\small, scale=0.82, transform shape
]
%% --- (a) Cavity schematic ---
\begin{scope}[yshift=.5cm]
  \node[cavity] (c1) at (0,0) {\textcolor{blue!50!black}{$U$}};
  \node[below=0.08cm of c1, font=\small] {site 1};
  \node[cavity] (c2) at (2.,0) {\textcolor{blue!50!black}{$U$}};
  \node[below=0.08cm of c2, font=\small] {site 2};
  % Hopping
\draw[{Stealth[length=5pt]}-{Stealth[length=5pt]}, thick, blue!60!black]
    (c1.east) to[bend left=-30] node[above, font=\small]{$J$} (c2.west);
  % Loss
  \draw[photon, green!60!black]
    ([xshift=0cm]c1.north) -- ++(0,1.)
    node[above, font=\footnotesize]{$\gamma$};
  \node[font=\footnotesize, green!60!black, align=center] at (0, 2.3) {classical\\light};
  \draw[photon, orange!90!black]
    ([xshift=0cm]c2.north) -- ++(0,1)
    node[above, font=\footnotesize]{$\gamma$};
  \node[font=\footnotesize, orange!90!black, align=center] at (2., 2.3) {quantum\\light};
  % CW drive on site 1
  \draw[photon, green!60!black, thick]
    (-1.5,0) -- (c1.west) node[midway, above, font=\footnotesize]{$F_1$};
  \node[font=\scriptsize, green!60!black] at (-1.0,-0.35) {CW};
  % Pulsed quadrature drive on site 2
  \draw[photon, orange!90!black, thick]
    (3.4,0) -- (c2.east) node[midway, above, font=\footnotesize]{$iF_1$};
  \node[font=\scriptsize, orange!90!black] at (3.4,-0.35) {CW/pulsed};
  \node at (-1.8, 4) {(a)};
\end{scope}

%% --- (b) Interference pathways ---
\begin{scope}[xshift=6.2cm, yshift=-1cm]
\end{scope}

\begin{scope}[xshift=6.0cm, yshift=0.0cm]
  % --- Levels ---
  \draw[thick] (-0.3,0) -- (0.3,0);
  \node[below, font=\scriptsize] at (0,0) {$|00\rangle$};
  \draw[thick] (-1.281,1.7) -- (-0.681,1.7);
  \node[left, font=\scriptsize] at (-1.281,1.7) {$|10\rangle$};
  \draw[thick] (0.681,1.7) -- (1.281,1.7);
  \node[right, font=\scriptsize] at (1.281,1.7) {$|01\rangle$};
  \draw[thick, dotted] (-2.262,3.4) -- (-1.662,3.4);
  \draw[thick] (-2.262,4.1) -- (-1.662,4.1);
  \node[left, font=\scriptsize] at (-2.262,4.1) {$|20\rangle$};
  \draw[thick] (-0.3,3.4) -- (0.3,3.4);
  \node[above, font=\scriptsize] at (0,3.4) {$|11\rangle$};
  \draw[thick, dotted] (1.662,3.4) -- (2.262,3.4);
  \draw[thick] (1.662,4.1) -- (2.262,4.1);
  \node[right, font=\scriptsize] at (2.262,4.1) {$|02\rangle$};
  % U spacing arrows
  \draw[{Stealth[length=4pt]}-{Stealth[length=4pt]}, thick]
    (-1.962,3.47) -- (-1.962,4.03)
    node[midway, right, font=\scriptsize]{$2U$};
  \draw[{Stealth[length=4pt]}-{Stealth[length=4pt]}, thick]
    (2.362,3.47) -- (2.362,4.03)
    node[midway, right, font=\scriptsize]{$2U$};
  % Crossed circle on middle of virtual level above |02>
  \draw[red!70!black, thick] (2.,4.1) circle (0.18cm);
  \draw[red!70!black, thick] (2.-0.127,4.1-0.127) -- (2.+0.127,4.1+0.127);
  \draw[red!70!black, thick] (2.-0.127,4.1+0.127) -- (2.+0.127,4.1-0.127);
  % --- Arrows ---
  % |00> -> |10>: goes left, green, F1
  \draw[arr, green!60!black, line width=1.2pt]
    (0,0.07) -- (-0.981,1.63)
    node[midway, left, font=\scriptsize, green!60!black]{$F_1$};
  % |00> -> |01>: goes right, orange, F2
  \draw[arr, orange!90!black, line width=1.2pt]
    (0,0.07) -- (0.981,1.63)
    node[midway, right, font=\scriptsize, orange!90!black]{$F_2$};
  % |10> <-> |01>: black bendy J
  \draw[{Stealth[length=4pt]}-{Stealth[length=4pt]}, blue!60!black, thick]
    (-0.581,1.7) to[bend right=30] (0.581,1.7);
  \node[font=\scriptsize] at (0,1.7) {$J$};
  % |10> -> |11>: goes right, orange, F2
  \draw[arr, orange!90!black, line width=1.2pt]
    (-0.981,1.77) -- (0,3.33)
    node[midway, right, font=\scriptsize, orange!90!black]{$F_2$};
  % |01> -> |11>: goes left, green, F1
  \draw[arr, green!60!black, line width=1.2pt]
    (0.981,1.77) -- (0,3.33)
    node[midway, left, font=\scriptsize, green!60!black]{$F_1$};
  % |10> -> virtual level below |20>: goes left, green, F1
  \draw[arr, green!60!black, line width=1.2pt]
    (-0.981,1.77) -- (-1.962,3.33)
    node[midway, left, font=\scriptsize, green!60!black]{$F_1$};
  % |01> -> virtual level below |02>: goes right, orange, F2
  \draw[arr, orange!90!black, line width=1.2pt]
    (0.981,1.77) -- (1.962,3.33)
    node[midway, right, font=\scriptsize, orange!90!black]{$F_2$};
  % |11> <-> |20>: black bendy J
  \draw[{Stealth[length=4pt]}-{Stealth[length=4pt]}, blue!60!black, thick]
    (-0.5,3.4) to[bend left=30] (-1.562,3.4);
  \node[font=\scriptsize] at (-1,3.4) {$J$};
  % |11> <-> |02>: black bendy J
  \draw[{Stealth[length=4pt]}-{Stealth[length=4pt]}, blue!60!black, thick]
    (0.5,3.4) to[bend right=30] (1.562,3.4);
  \node[font=\scriptsize] at (1,3.4) {$J$};
  % Panel label
  \node at (-3.0,4.5) {(b)};
\end{scope}
\end{tikzpicture}
\caption{(a)~Kerr dimer setup: site~1 driven by a CW field $F_1$,
site~2 by a pulsed quadrature field $iF_1$; both sites decay at rate $\gamma$
and are coupled by hopping $J$.
(b)~Two-photon energy levels and transition amplitudes.
The two-photon occupation of site~2, $|0,2\rangle$, receives contributions
from direct two-photon excitation via $F_2$ and from indirect transfer
via the $|1,1\rangle$ intermediate state and hopping $J$.
These two channels interfere destructively at the UPB condition,
giving zero two-photon occupation of site~2 (crossed circle).}
\label{fig:schematic}
\end{figure}
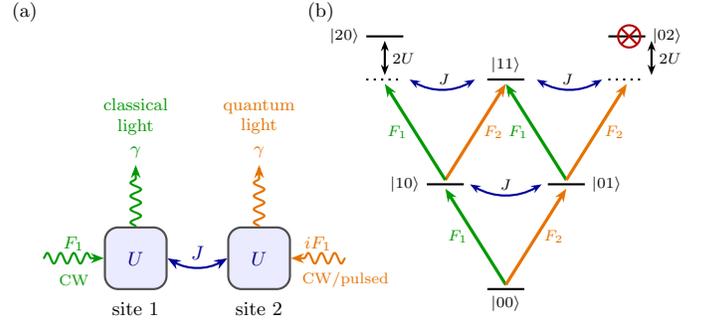

Generating single photons on demand is central to quantum communication
and quantum information processing~\cite{couteau_applications_2023}.  Conventional photon blockade suppresses
multi-photon occupation of a resonator by requiring the Kerr nonlinearity
$U$ to exceed the cavity decay rate $\gamma$
\cite{tian_quantum_1992,imamoglu_strongly_1997}, a phenomenon first observed in
cavity QED \cite{birnbaum_photon_2005} and subsequently demonstrated in circuit QED
\cite{lang_observation_2011}, and on-chip photonic
systems \cite{faraon_coherent_2008}.  In practice, the large nonlinearity required
restricts conventional photon blockade to discrete emitters --- chiefly
self-assembled semiconductor quantum dots \cite{michler_quantum_2000,senellart_high-performance_2017}
and atoms \cite{birnbaum_photon_2005} --- where the quantised energy spectrum provides
the necessary anharmonicity. Self-assembled quantum dots require individual
post-fabrication tuning for indistinguishable multi-source emission
\cite{senellart_high-performance_2017}, limiting scalability.  Extending antibunched light
generation to continuous-medium platforms such as quantum-well or 2D-material
polaritons \cite{delteil_towards_2019,munoz-matutano_emergence_2019,zhang_van_2021} --- where many
identical sources can be etched in a single lithographic step --- would be
transformative, but material nonlinearities in these systems are orders of
magnitude below cavity linewidths, placing conventional blockade out of reach.

Unconventional photon blockade (UPB) \cite{liew_single_2010,bamba_origin_2011,ferretti_photon_2010}
side-steps this requirement by exploiting destructive quantum interference
between two-photon pathways in a coupled cavity dimer, achieving
$g^{(2)}(0) \approx 0$ with $U \ll \gamma$. A key limitation is that
the interference requires strong inter-cavity coupling $J \gg \gamma$, inducing
rapid oscillations in $g^{(2)}(\tau)$ at frequency $2J$, with an antibunching
window $\delta\tau \sim 1/J \ll 1/\gamma$ \cite{flayac_unconventional_2017}.  At
telecom wavelengths this window is only $\sim 100\,\text{ps}$
\cite{flayac_all-silicon_2015}, at the edge of detector resolution.  Compounding this, the
condition $J \gg \gamma$ is itself challenging in photonic systems: a large
coupling with a small loss rate requires extremely high-$Q$ resonators with
tight mode overlap, difficult to reconcile with the fabrication tolerances
needed for fine-tuned UPB parameters.

% Two recent proposals address these constraints.
% Flayac and Savona \cite{flayac_single_2016} proposed replacing the inter-cavity 
% hopping with a one-way dissipative coupling to suppress the rapid oscillations
% in $g^{(2)}(\tau)$, though such non-reciprocal couplings are experimentally
% challenging to realise. Wang et al.\ \cite{wang_long-lived_2025} engineered cavity networks with
% single-photon dark states at large $\gamma/J$, achieving long-lived photon
% blockade with $g^{(2)}(\tau) \propto (\gamma\tau)^4$ over a window of
% $\sim 8/\gamma$.  Here we
% pursue a complementary approach that retains the minimal two-cavity geometry:
% we replace the single-site drive with a bilateral drive $F_2 = F_1 e^{i\phi}$
% of \emph{equal amplitude} and ask which phase $\phi$ satisfies UPB
% (Fig.~\ref{fig:schematic}a).  We show that the quadrature choice
% $\phi = 90^\circ$ reduces $J_{\min}$ by a factor of $2\sqrt{2}$, gives an
% analytic parameter locus, and places the system naturally in the overdamped
% regime $J < \gamma/2$ --- eliminating rapid oscillations in $g^{(2)}(\tau)$.
% We further demonstrate that the scheme operates under pulsed excitation,
% enabling on-demand single-photon generation.

Two recent proposals address these constraints.
Flayac and Savona \cite{flayac_single_2016} proposed replacing the inter-cavity
hopping with a one-way dissipative coupling to suppress the rapid oscillations
in $g^{(2)}(\tau)$.  While nonreciprocal photonic elements have been
demonstrated --- for instance, asymmetric backscattering between
counter-propagating modes near an exceptional point \cite{peng_chiral_2016} ---
the cascaded, unidirectional dissipative coupling between two separate cavities
that this scheme requires remains a distinct and demanding requirement. Wang et al.\ \cite{wang_long-lived_2025} engineered cavity networks with
single-photon dark states at large $\gamma/J$, achieving long-lived photon
blockade with $g^{(2)}(\tau) \propto (\gamma\tau)^4$ over a window of
$\sim 8/\gamma$.

Bilateral driving of a Kerr dimer with two coherent fields of tunable
relative phase has itself been studied as a route to UPB
\cite{shen_exact_2015,wang_unconventional_2017,shen_unconventional_2018}, with the phase
and amplitude ratio scanned to optimise antibunching; a related input--output
approach instead mixes the two cavity outputs on a beamsplitter to isolate the
antibunched mode \cite{flayac_input-output_2013,flayac_unconventional_2017}, while an emitter-based scheme
unifies conventional and unconventional blockade in a two-photon
Jaynes--Cummings model \cite{zhou_universal_2025}.  Here we retain the minimal
two-cavity geometry but single out the equal-amplitude quadrature drive
$F_2 = iF_1$, which produces antibunching directly on a single physical
port (site~2) with no output mixing and no nonreciprocal coupling
(Fig.~\ref{fig:schematic}a).  For this drive the coupling threshold is
$J_{\min}=\gamma/4$ --- a factor of $2\sqrt{2}$ below the single-site value
$\gamma/\sqrt{2}$ \cite{bamba_origin_2011} --- and the entire band
$\gamma/4 < J < \gamma/\sqrt{2}$, which admits no single-site UPB
solution for any $U$, becomes accessible, including the overdamped regime
$J<\gamma/2$ where $g^{(2)}(\tau)$ rises smoothly without the $2J$ oscillations.
We give the parameter locus in closed form, show that on-demand pulsed
operation is possible without tailoring the pulse, and that fabrication
disorder can be fully compensated by re-tuning the drive.

\paragraph{Model.}
Two coupled single-mode Kerr cavities are driven coherently under the
Hamiltonian (rotating frame, $\hbar = 1$)
\begin{align}
  \hat{H} &= \Delta(\hat{n}_1+\hat{n}_2)
  + U\!\sum_j\hat{a}_j^{\dagger2}\hat{a}_j^2
  + J(\hat{a}_1^\dagger\hat{a}_2+\hat{a}_2^\dagger\hat{a}_1) \nonumber\\
  &\quad + F_1(\hat{a}_1^\dagger+\hat{a}_1)
  + (F_2\hat{a}_2^\dagger+F_2^*\hat{a}_2),
  \label{eq:H}
\end{align}
where $\Delta$ is the cavity-laser detuning and $U$ the on-site Kerr
nonlinearity.  Each site decays at rate $\gamma$ via Lindblad photon loss.
We set $F_1\in\mathbb{R}$, $F_2 = F_1 e^{i\phi}$, and
$\tilde{E} \equiv \Delta - i\gamma/2$.

In the weak-drive limit $F_1\ll\gamma$, quantum jump terms are suppressed
by $F_1/\gamma$ relative to the no-jump terms at each order in the Fock
expansion (Appendix~\ref{sec:derivation}), so the open-system
dynamics reduces to a non-Hermitian Schr\"{o}dinger equation.  The steady
state is therefore well approximated by a Fock-state amplitude expansion
truncated at two photons \cite{bamba_origin_2011},
$|\psi_\text{ss}\rangle \approx |0,0\rangle + \sum_{m+n\leq2}C_{mn}|m,n\rangle$,
with $g^{(2)}_{22}(0) = 2|C_{02}|^2/|C_{01}|^4$.
UPB on site~2 requires $C_{02} = 0$.

\paragraph{Analytic results.}
With $p\equiv e^{i\phi}$, the one-photon equations give $C_{01}=rC_{10}$
with $r = (J-p\tilde{E})/(pJ-\tilde{E})$ (see Appendix~\ref{sec:derivation}).
For $p=i$ ($\phi=90^\circ$), $|r|<1$ for all $J,\Delta,\gamma>0$:
site~2 is always less occupied than site~1.
Imposing $C_{02}=0$ in the two-photon sector, we derive the exact
condition
\begin{equation}
  J^2\bigl(2\tilde{E}+U(1+p^2)\bigr)
  = 2p\tilde{E}(\tilde{E}+U)(2J-p\tilde{E}),
  \label{eq:general}
\end{equation}
solved explicitly for $U$ as
\begin{equation}
  U = \frac{-2\tilde{E}(J-p\tilde{E})^2}
           {J^2(1+p^2)-4pJ\tilde{E}+2p^2\tilde{E}^2}.
  \label{eq:U_phi}
\end{equation}
The physical solution (real, positive $U$) requires $\mathrm{Im}(U)=0$,
determining $\Delta_\mathrm{opt}(\phi,J)$.  No solution exists at $\phi=0$
or $\pi$: at these phases the two indirect pathways add constructively rather
than destructively, precluding cancellation of $C_{02}$ (Appendix~\ref{sec:general_phi}).  For $J>\gamma/2$, a $U=0$ solution appears at
$\phi=\arcsin(\gamma/2J)$, but this is a \emph{linear dark state} with
$n_2=0$ identically — not genuine UPB (Appendix~\ref{sec:dark_state}).

\begin{figure}[t]
\centering
\includegraphics[width=\columnwidth]{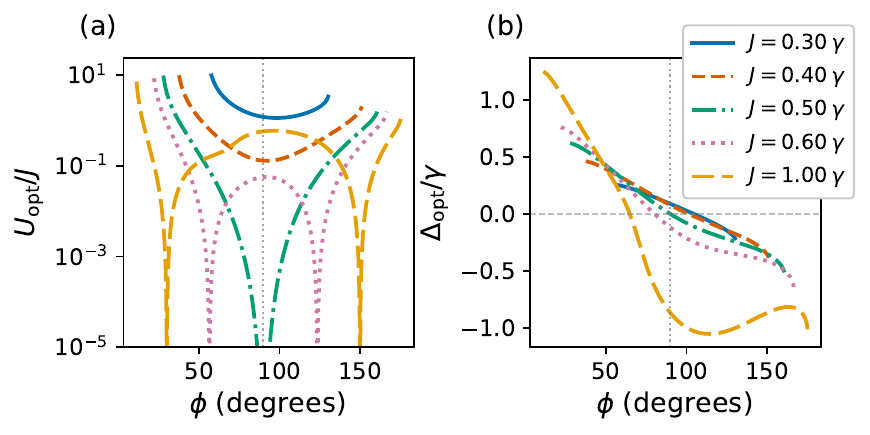}
\caption{UPB parameter locus vs drive phase $\phi$.
\textit{Left}: required nonlinearity $U_\mathrm{opt}/J$ (log scale).
\textit{Right}: optimal detuning $\Delta_\mathrm{opt}/\gamma$.
Four values of $J/\gamma$; dotted vertical line: $\phi=90^\circ$.}
\label{fig:phi_locus}
\end{figure}

Setting $p=i$, the factor $1+p^2$ vanishes and \eqref{eq:general} collapses
to the compact quadratic
\begin{equation}
  (\tilde{E}+U)(\tilde{E}+2iJ) = J^2,
  \label{eq:main}
\end{equation}
whose real and imaginary parts give the closed-form parameter locus
\begin{align}
  \Delta_\mathrm{opt} &= \Bigl(\tfrac{\gamma}{2}-J\Bigr)
                         \sqrt{\tfrac{4J}{\gamma}-1},
  \label{eq:Delta}\\[3pt]
  U_\mathrm{opt}      &= \frac{4(J-\gamma/2)^2}{\sqrt{4J/\gamma-1}},
  \label{eq:U}
\end{align}
valid for $J>J_\mathrm{min}=\gamma/4$ — a factor of $2\sqrt{2}\approx2.83$
below the single-site pumped UPB threshold $\gamma/\sqrt{2}$ \cite{bamba_origin_2011}.

The physical content of Eq.~\eqref{eq:main} is that the direct drive
$F_2|0,1\rangle\!\to\!|0,2\rangle$ is exactly cancelled by hopping from
$|1,1\rangle$ [Fig.~\ref{fig:schematic}(b)], regardless of how $|1,1\rangle$
was populated.  The Kerr shift $U$ detunes the two-photon resonance,
providing the degree of freedom needed for this cancellation to occur at
a real detuning $\Delta$; the master condition \eqref{eq:main} is the
self-consistency requirement that the amplitude of $|1,1\rangle$ is
consistent with its equation of motion
(Appendix~\ref{sec:derivation}).

\begin{figure}[t]
\vspace{0.15cm}
\centering
\includegraphics[width=\columnwidth]{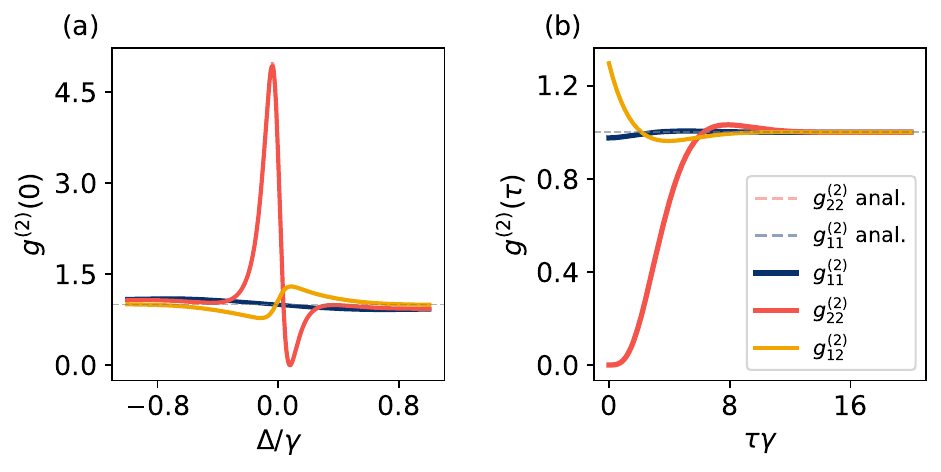}
\caption{Second-order correlators at the optimal point
($J=0.4\gamma$, $U=0.052\gamma$, $\Delta=0.077\gamma$). Solid curves are full Lindblad master-equation simulations, dashed curves the analytic
results (indistinguishable on this scale).
(a)~Equal-time correlators $g^{(2)}_{11}(0)$ (navy), $g^{(2)}_{22}(0)$ (red),
and $g^{(2)}_{12}(0)$ (yellow) vs detuning $\Delta$ at fixed $F_1=0.01\gamma$;
dashed curves are the analytic results (indistinguishable from the numerics on this scale).
(b)~Time-delayed correlators $g^{(2)}_{11}(\tau)$ (navy),
$g^{(2)}_{22}(\tau)$ (red), and $g^{(2)}_{12}(\tau)$ (yellow);
dashed curves are the analytic results (indistinguishable from the numerics on this scale).}
\label{fig:g2tau}
\end{figure}

Solutions exist only in a range of $\phi$ that broadens with increasing
$J$ but is not symmetric around $90^\circ$, and $U_\mathrm{opt}$ is
minimised close to $\phi=90^\circ$ (within $0.7\%$ of the global minimum
for $J=0.4\gamma$), making the quadrature drive the natural operating point
when the Kerr interaction is most constrained
(Fig.~\ref{fig:phi_locus}).
Figure~\ref{fig:g2tau}(a) shows how all three equal-time correlators vary with
detuning $\Delta$ at the UPB optimum.
$g^{(2)}_{11}(0)$ remains close to unity throughout, confirming that site~1
retains near-Poissonian statistics.
$g^{(2)}_{22}(0)$ exhibits strong bunching for detunings below the optimum —
reflecting two-photon resonance enhancement — before dropping sharply to
zero at $\Delta_\mathrm{opt}$, where the interference condition is satisfied.

\begin{figure*}[t]
\centering
\includegraphics[width=\textwidth]{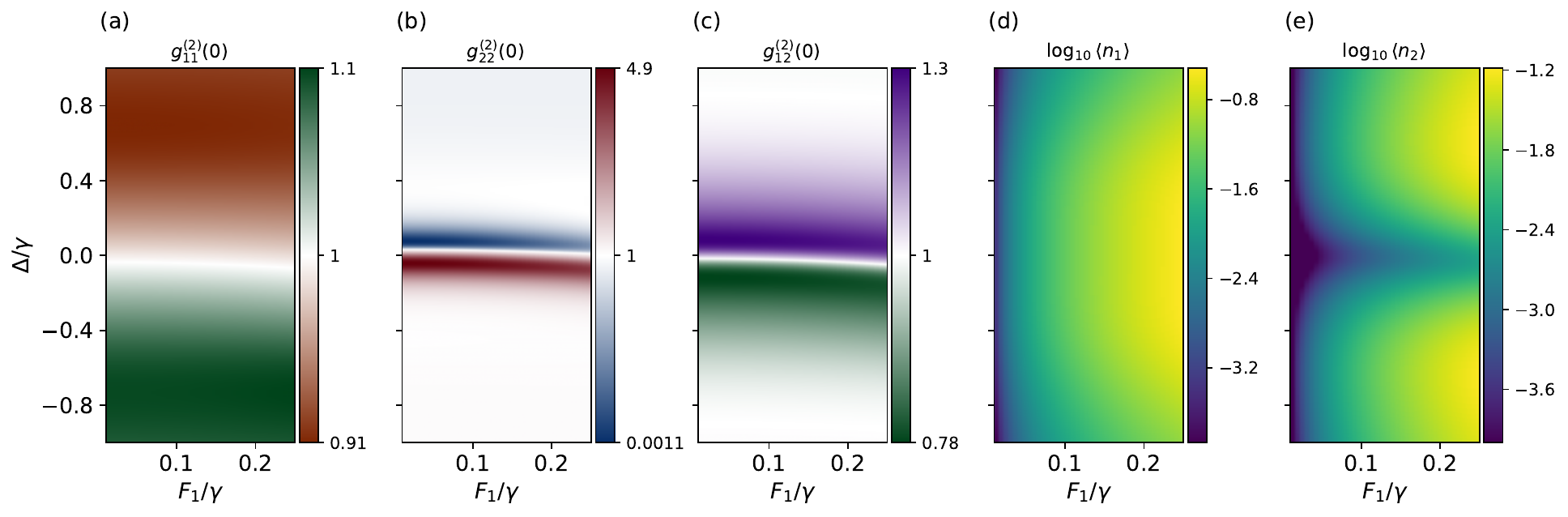}
\caption{Parameter landscape at $\phi=90^\circ$, $J=0.4\gamma$, $U=0.052\gamma$,
as a function of detuning $\Delta/\gamma$ (vertical) and drive amplitude
$F_1/\gamma\in[0.01,\,0.25]$ (horizontal). All panels are full Lindblad master-equation simulations.
Panels from left: $g^{(2)}_{11}(0)$, $g^{(2)}_{22}(0)$, $g^{(2)}_{12}(0)$,
$\log_{10}\langle n_1\rangle$, $\log_{10}\langle n_2\rangle$.}
\label{fig:landscape}
\end{figure*}

\paragraph{Overdamped regime and time-delayed correlator.}
The threshold $J_\mathrm{min}=\gamma/4$ places the natural operating point at
$J\lesssim\gamma/2$, the overdamped regime where $\gamma$ exceeds the
inter-mode oscillation frequency $2J$. The suppression of the rapid oscillations is therefore indirect: the
quadrature drive does not itself modify the normal-mode dynamics, but lowers
the UPB threshold from $\gamma/\sqrt{2}$ to $\gamma/4$, permitting operation
at $J<\gamma/2$, where the beat frequency $2J$ falls below the decay rate
$\gamma$. We derive a closed-form expression for
$g^{(2)}_{22}(\tau)$ via the quantum regression theorem in the weak-drive
limit \cite{shen_exact_2015} (Appendix~\ref{sec:g2_analytic}).
After detecting a photon from site~2, the post-detection state
$\hat{a}_2|\psi_\mathrm{ss}\rangle = C_{01}|0,0\rangle + C_{11}|1,0\rangle$
(using $C_{02}=0$) lies entirely in the one-photon sector.
It evolves under the driven one-photon equations, which diagonalise into
two decoupled modes with decay rates $u_{1,2} = \gamma/2 \mp i\omega_{1,2}$
and oscillation frequencies $\omega_{1,2} = \Delta_\mathrm{opt} \pm J$. The real part $\gamma/2$ is common to both normal modes and independent of $J$:
the coherent coupling shifts only the mode frequencies (splitting them by
$2J$) and introduces no additional loss.  ``Overdamped'' here denotes
$2J<\gamma$, i.e.\ the inter-mode beat is damped before completing an
oscillation, in the sense of a damped harmonic oscillator.
The result $g^{(2)}_{22}(\tau) = |C_{01}(\tau)|^2/|C_{01}|^4$ satisfies
$g^{(2)}_{22}(0)=0$ and $g^{(2)}_{22}(\infty)=1$ exactly; the analytic
$g^{(2)}_{11}(\tau)$ and $g^{(2)}_{22}(\tau)$ are shown in
Fig.~\ref{fig:g2tau}(b), where $g^{(2)}_{11}(\tau)$ remains close to unity
at all delays while $g^{(2)}_{22}(\tau)$ rises smoothly from zero without
rapid oscillations, crossing $0.5$ at $\tau\approx3.2/\gamma$.

% The smooth, monotone rise of $g^{(2)}_{22}(\tau)$ to unity is confirmed
% numerically.  The peak overshoot $g^{(2)}_\mathrm{max}-1$ is
% non-monotonic across $J$: it grows from $0.4\%$ near threshold to $8\%$
% near $J=\gamma/2$, then decreases in the underdamped regime, reaching
% $0.04\%$ at $J=\gamma$ despite oscillation periods $T_2\sim 3/\gamma$.

The smooth rise of $g^{(2)}_{22}(\tau)$ to unity, free of rapid
oscillations, is confirmed numerically.  Quantitatively, by
``oscillation-free'' we mean that in the overdamped regime the transient
consists of at most a single overshoot of unity, with no multi-peaked
ringing at frequency $2J$; genuine oscillations appear only for
$J>\gamma/2$.  The peak overshoot $g^{(2)}_\mathrm{max}-1$ is
non-monotonic across $J$: it grows from $0.4\%$ near threshold to $8\%$
near $J=\gamma/2$ ($3.3\%$ at the operating point $J=0.4\gamma$), then
decreases in the underdamped regime, reaching $0.04\%$ at $J=\gamma$
despite oscillation periods $T_2\sim 3/\gamma$. The shrinking overshoot reflects decreasing transient amplitude relative to
the steady-state value $|C_{01}|^2$ (Appendix~\ref{sec:osc_amplitude} gives the full $J$-scan and tabulated values).

To see how the correlators behave at higher drive amplitudes and across
a range of detunings, we solve the full master equation numerically (see Appendix~\ref{sec:numerics}).
At $F_1=0.01\gamma$ the simulations match the analytic results nearly
perfectly (Fig.~\ref{fig:g2tau});  the breakdown of the no-jump approximation with increasing $F_1$ is quantified in Appendix~\ref{sec:numerics}
(see also Fig.~\ref{fig:landscape_cuts}), where the minimum of
$g^{(2)}_{22}(0)$ is seen to rise and redshift as $F_1$ increases.
The full dependence on $F_1$ and $\Delta$ is shown in
Fig.~\ref{fig:landscape}: $g^{(2)}_{11}(0)$ hovers close to unity
throughout the landscape, while at the maximum drive $F_1=0.25\gamma$
the minimum $g^{(2)}_{22}(0)$ rises to $\approx0.46$, with mean
occupations $\langle n_1\rangle\approx0.30$ and
$\langle n_2\rangle\approx7\times10^{-3}$.

\begin{figure}[t]
\centering
\includegraphics[width=\columnwidth]{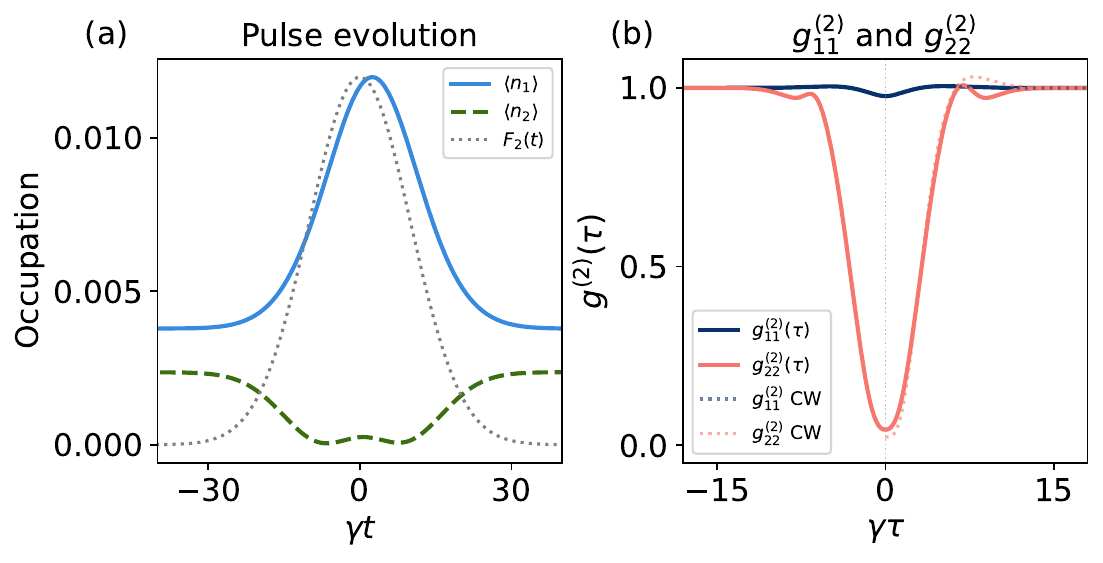}
\caption{Pulsed excitation ($J=0.4\gamma$, $\sigma=10/\gamma$, $F=0.05\gamma$).
\textit{Left}: occupations $\langle n_1\rangle$ (blue), $\langle n_2\rangle$
(dashed green), and scaled pulse envelope (dotted).
\textit{Right}: $g^{(2)}_{11}(\tau)$ (navy) and $g^{(2)}_{22}(\tau)$
(red), with CW references (dotted). $\tau=0$ is referenced to the pulse peak centre ($t=0$).}
\label{fig:pulsed}
\end{figure}

\paragraph{Pulsed excitation.}
On-demand single-photon emission requires pulsed operation. Earlier UPB
single-photon proposals operated under pulsed excitation either within the
narrow $\delta\tau\sim1/J$ antibunching window~\cite{flayac_all-silicon_2015}
or by removing the normal-mode oscillations through a nonreciprocal one-way
dissipative coupling~\cite{flayac_single_2016}.  The overdamped regime
accessed here achieves the same clean pulsed operation on a standard
Hermitian dimer: the $\sim1/\gamma$ antibunching window means the
antibunching survives pulsed excitation without overshoots or oscillations.
We simulate a hybrid drive: site~1 is pumped continuously at amplitude $F_1$
while site~2 receives a Gaussian pulse
$F_2(t)=iF_1\,e^{-t^2/2\sigma^2}$ with $F_1=0.05\gamma$.  This
models a CW pump that maintains a steady population in the dimer, with a
pulsed quadrature field triggering single-photon emission from site~2.
For $\sigma=10/\gamma$, the equal-time value $g^{(2)}_{22}(0)$ at the pulse
peak approaches the CW dip, confirming that the antibunching is robust on
the timescale of the pulse width and free of ringing
(Fig.~\ref{fig:pulsed}).
For a GaAs quantum-well photonic molecule with $Q\simeq10^4$ at
$\lambda=810\,\text{nm}$, $1/\gamma\approx4.3\,\text{ps}$ and the pulse
width $\sigma=10/\gamma\approx43\,\text{ps}$ gives a repetition rate of
$\sim5\,\text{GHz}$.  At $F_1=0.05\gamma$ the mean
occupation reaches $\langle n_2\rangle\approx0.012$, giving a detected
photon rate of $\sim20\,\text{MHz}$ with $g^{(2)}_{22}(0)<0.05$ ---
competitive with state-of-the-art GaAs single-photon sources \cite{senellart_high-performance_2017}. Because the $\sim1/\gamma$ antibunching window spans essentially the whole
emission, this rate requires no detector time-gating; in conventional UPB the 
$\sim1/J$ window forces gating onto a small fraction of the emission, reducing
the usable rate accordingly.

\paragraph{Discussion.}
The bilateral quadrature drive achieves two decisive advantages over the
single-site-driven UPB scheme \cite{bamba_origin_2011}.
First, the single-site scheme requires $J>\gamma/\sqrt{2}$ for any solution
to exist; below this threshold no value of $U$ can satisfy the interference
condition.  Our scheme operates for $J>\gamma/4$, opening the entire
parameter space $\gamma/4 < J < \gamma/\sqrt{2}$ that is inaccessible to
the single-site drive, including the overdamped regime $J<\gamma/2$ where
$g^{(2)}(\tau)$ is smooth and oscillation-free.
At the operating point $J=0.4\gamma$ we obtain $U\approx0.052\gamma$
with no counterpart in the single-site scheme.
For $J \gtrsim 0.96\,\gamma$ the single-site $U$
eventually falls below ours (the two are equal at $J\approx0.96\,\gamma$,
$U\approx0.50\,\gamma$), but that regime requires $J\gg\gamma$,
placing extreme demands on cavity $Q$-factors and yielding a rapidly
oscillating $g^{(2)}(\tau)$ with antibunching window $\delta\tau\sim 1/J$.
A minimal two-cavity geometry driven with a single laser and a
$90^\circ$ electro-optic phase shift is therefore sufficient to achieve UPB
in the experimentally favourable overdamped regime.

The use of a controlled inter-port phase to enforce destructive interference in a two-mode system recalls coherent perfect absorption \cite{chong_coherent_2010, zanotto_perfect_2014}, where a tuned relative phase between two input ports suppresses the linear response; here the same principle is applied one level higher in the Fock hierarchy, selectively cancelling the two-photon amplitude while leaving the one-photon output intact.

The general condition \eqref{eq:U_phi} shows that
$\phi=90^\circ$ minimises $U_\mathrm{opt}$ to within $0.7\%$ of the global
optimum for $J<\gamma/2$, making it the natural operating point when Kerr
interaction strength is the limiting constraint.  Along the locus
\eqref{eq:Delta}--\eqref{eq:U}, $U_\mathrm{opt}$ has no interior minimum
with $U>0$: it decreases monotonically toward zero as $J\to(\gamma/2)^-$
(the linear dark-state boundary) and increases for $J>\gamma/2$.  The
operating point $J\approx0.4\gamma$, giving $U_\mathrm{opt}\approx0.052\gamma$,
balances a small required nonlinearity against sufficient distance from this
boundary.  Promising platforms include
Cu$_2$O Rydberg exciton--polaritons \cite{orfanakis_rydberg_2022}, semiconductor
photonic molecules \cite{galbiati_polariton_2012, abbarchi_macroscopic_2013}, trion polaritons in 2D monolayers
\cite{kyriienko_nonlinear_2020}, and Moir\'e polariton systems
\cite{zhang_van_2021} where enhanced nonlinearities are accessible.

A particularly compact implementation is possible within a \emph{single}
birefringent micropillar~\cite{reitzenstein_polarization-dependent_2010}, where the H and V polarisation modes play the
roles of the two sites, the polarisation splitting $\delta$ provides the
hopping $J=\delta/2$, and the quadrature drive condition $F_2=iF_1$ is
satisfied by circularly polarised excitation --- requiring only a
quarter-wave plate in the beam path.  The single-pillar geometry
introduces an additional cross-Kerr interaction $U_x\hat{n}_H\hat{n}_V$
from opposite-spin exciton--exciton scattering, which one might worry
could perturb the UPB condition. However, since $U_x$ shifts only the $|1,1\rangle$ two-photon state ---
leaving $|0,2\rangle$ and $|2,0\rangle$ unchanged --- it enters the UPB
condition only at first order through the self-consistency of $C_{11}$.
For GaAs, where the opposite-spin interaction $U_x$ is small and
negative \cite{vladimirova_polariton-polariton_2010}, $U_\mathrm{opt}$ shifts by
less than $3\%$ across the full experimentally reported range of $U_x/U$,
while $\Delta_\mathrm{opt}$ may shift by up to $\sim10\%$ at the largest
values,
and the minimum coupling threshold $J_\mathrm{min}=\gamma/4$ is unchanged
to all orders in $U_x/U$.  The single-pillar scheme is therefore as robust
as the two-site dimer.

We quantify the sensitivity to fabrication disorder
and its compensation in Appendix~\ref{sec:disorder}:
with fixed drive parameters the $g^{(2)}_{22}(0)<0.1$ condition is
maintained for detuning mismatches up to $\pm0.033\,\gamma$
($\sim1\,\text{GHz}$ in GaAs), but re-tuning the drive phase $\phi$
alone extends this tolerance nearly tenfold to $\pm0.26\,\gamma$, and
adjusting both $\phi$ and the amplitude ratio $|F_2|/|F_1|$ restores
exact UPB, eliminating the need for
post-fabrication cavity trimming. Multiple identical dimers can be 
etched across the same semiconductor wafer in a single lithographic step, with low site-to-site disorder and
near-identical emission frequencies, offering a route to parallelised
single-photon generation with high photon indistinguishability that is
difficult to achieve with self-assembled quantum dots without
post-fabrication tuning.

The dimer unit tiles naturally into a driven-dissipative lattice of
symmetrically driven Kerr dimers, where the overdamped condition
$J<\gamma/2$ relaxes quality-factor requirements compared to conventional
UPB schemes, providing a practical starting point for exploring quantum
many-body phases in weakly nonlinear photonic and polaritonic arrays.

%maybe?
% We note that, like all weak-drive UPB schemes~\cite{lemonde_antibunching_2014},
% the emitted field is a Gaussian amplitude-squeezed state; the small
% $g^{(2)}_{22}(0)$ reflects suppressed two-photon occupation rather than a
% Fock-state single photon.  For applications requiring non-Gaussian resources
% this is a fundamental limitation, though for quantum key distribution and
% linear-optical schemes sub-Poissonian statistics suffice.

In conclusion, equal-amplitude bilateral driving with a $90^\circ$ phase difference
reduces the UPB coupling threshold to $J_\mathrm{min}=\gamma/4$ and places
the system in the overdamped regime, yielding a smooth $g^{(2)}(\tau)$
directly resolvable with standard detectors.  Unlike schemes requiring
non-reciprocal couplings or large photon networks, the approach uses only
a single laser with a controlled phase shift and a standard Hermitian dimer,
well suited to weakly nonlinear platforms such as Cu$_2$O Rydberg
exciton--polaritons \cite{orfanakis_rydberg_2022} and semiconductor photonic
molecules \cite{galbiati_polariton_2012} where conventional photon blockade is beyond reach. 
Site-to-site fabrication disorder can be compensated by re-tuning the drive:
adjusting the phase alone extends the disorder tolerance nearly tenfold, and
adjusting both the phase and the amplitude ratio restores the blockade
exactly, removing the need for post-fabrication cavity trimming and opening
a route to arrays of identical single-photon sources on a single chip.  The minimal two-cavity
geometry and overdamped operating condition make the scheme directly
extensible to driven-dissipative lattices, providing a scalable platform
for quantum many-body physics in weakly nonlinear photonic and polaritonic
systems.

\paragraph{Data availability.}
The simulation scripts, numerical output data, and plotting code used to produce all figures in this manuscript are available in the University of St Andrews Research Portal at \url{https://doi.org/10.17630/27f9c1cf-e543-40d1-a207-13c69ce2b9fb}.

\begin{acknowledgments}
I thank T.~C.~H.~Liew and J.~M.~J.~Keeling for stimulating discussions.
I acknowledge the use of Claude (Anthropic) for assistance in literature review, manuscript preparation, simulation code and calculations.
\end{acknowledgments}

\appendix

% ============================================================
\section{Derivation of the $C_{02}=0$ Condition at $\phi=90^\circ$}
\label{sec:derivation}
% ============================================================

\subsection{Hamiltonian and Lindblad Master Equation}

We consider two coupled single-mode Kerr cavities driven coherently in the
rotating frame ($\hbar=1$), with Hamiltonian as in the main text
(Eq.~\ref{eq:H}).  Photon loss from each site at rate $\gamma$ is modelled by
the Lindblad master equation:
\begin{equation}
    \dot{\rho} = -i[\hat{H}, \rho]
    + \gamma\Bigl(
    \sum_{k=1,2}\hat{a}_k\rho\hat{a}_k^\dagger
    - \tfrac{1}{2}\{\hat{N},\rho\}
    \Bigr),
    \label{eq:lindblad}
\end{equation}
where $\hat{N}=\hat{n}_1+\hat{n}_2$ is the total photon number operator,
valid since both sites share the same decay rate $\gamma$.
We use the \emph{Kerr convention} $U\hat{a}^{\dagger 2}\hat{a}^2$ without
an additional factor of $1/2$, matching Ref.~\cite{bamba_origin_2011},
so the two-photon state $|2\rangle$ acquires an energy shift $2U$ relative
to the non-interacting case.

\subsection{No-Jump Limit and Weak-Drive Fock-State Expansion}

In the weak-drive limit $F_1\ll\gamma$, density matrix elements scale as
$\rho_{mn,m'n'}\sim(F_1/\gamma)^{(m+n+m'+n')/2}$, since each photon costs
one drive insertion suppressed by $1/\gamma$.  The jump superoperator
contributes to the equation of motion for $\rho_{mn,m'n'}$ via the source
term
\begin{equation}
  \gamma\sum_k\langle m,n|\hat{a}_k\rho\hat{a}_k^\dagger|m',n'\rangle
  \propto \rho_{m+1,n;\,m'+1,n'} + \rho_{m,n+1;\,m',n'+1},
\end{equation}
which samples $\rho$ at total photon number $m+n+1$ (or $m'+n'+1$) ---
one order higher in $F_1/\gamma$ than $\rho_{mn,m'n'}$ itself.  The jump
contribution to the amplitude equations is therefore suppressed by
$F_1/\gamma$ relative to the no-jump terms and can be neglected throughout
the truncated space.  The open-system dynamics reduces to a non-Hermitian
Schr\"{o}dinger equation governed by the effective Hamiltonian
\begin{equation}
    \hat{H}_\mathrm{eff} = \hat{H} - i\tfrac{\gamma}{2}\hat{N},
    \label{eq:Heff}
\end{equation}
where $\hat{N}=\hat{n}_1+\hat{n}_2$.  This replaces the real cavity
detuning $\Delta$ with the complex frequency
$\tilde{E}\equiv\Delta-i\gamma/2$ throughout all amplitude equations,
encoding the cavity decay as an imaginary part of the single-photon energy.

The steady state is therefore well approximated by the pure-state ansatz
\begin{equation}
    |\psi_\mathrm{ss}\rangle \simeq
    \sum_{n+m \leq 2} C_{nm}|n,m\rangle,
\end{equation}
with $C_{00}\approx1$.  The correlators are
$g^{(2)}_{11}(0)=2|C_{20}|^2/|C_{10}|^4$ and
$g^{(2)}_{22}(0)=2|C_{02}|^2/|C_{01}|^4$; UPB on site~2
requires $C_{02}=0$.

\subsection{One-Photon Sector: Exact Symmetry for $F_2=iF_1$}

With the amplitude equations governed by $\hat{H}_\mathrm{eff}$
[Eq.~\eqref{eq:Heff}], projecting onto $|1,0\rangle$ and $|0,1\rangle$ and setting
$\dot{C}_{10}=\dot{C}_{01}=0$ gives the linear system
\begin{equation}
    \begin{pmatrix} \tilde{E} & J \\ J & \tilde{E} \end{pmatrix}
    \begin{pmatrix} C_{10} \\ C_{01} \end{pmatrix}
    = \begin{pmatrix} -F_1 \\ -F_2 \end{pmatrix},
    \label{eq:1ph_system}
\end{equation}
where $\tilde{E}=\Delta-i\gamma/2$ appears on both diagonal entries since
both sites share the same detuning and decay rate.  For $F_2=iF_1$ one verifies
immediately that $C_{01}=rC_{10}$ with
\begin{equation}
    r = \frac{i\tilde{E}-J}{\tilde{E}-iJ}.
    \label{eq:r}
\end{equation}
Since $(J+\gamma/2)^2>(J-\gamma/2)^2$ for all $J,\gamma>0$, we have
$|r|<1$ unconditionally: site~2 is always less occupied than site~1
despite $|F_2|=|F_1|$.

\subsection{Two-Photon Sector and the UPB Condition}

Projecting onto $|2,0\rangle$, $|1,1\rangle$, $|0,2\rangle$ gives three
coupled equations.  Defining $E_U\equiv\tilde{E}+U$ and using the
\emph{crucial identity} $E_2\equiv 2\Delta-i\gamma=2\tilde{E}$ (valid for
identical cavities), the steady-state equations are:
\begin{align}
    2E_U\,C_{20}+\sqrt{2}J\,C_{11} &= -\sqrt{2}F_1 C_{10},
    \label{eq:A}\\
    \sqrt{2}J\,C_{20}+2\tilde{E}\,C_{11}+\sqrt{2}J\,C_{02}
    &= -F_2 C_{10}-F_1 C_{01},
    \label{eq:B}\\
    \sqrt{2}J\,C_{11}+2E_U\,C_{02} &= -\sqrt{2}F_2 C_{01}.
    \label{eq:C}
\end{align}
Setting $C_{02}=0$ in~\eqref{eq:C} gives
$C_{11}=-iF_1 C_{01}/J$.  This has a direct physical meaning: the
steady-state equation for $C_{02}$,
\begin{equation}
    2(\tilde{E}+U)\,C_{02} + \sqrt{2}J\,C_{11} = -\sqrt{2}F_2\,C_{01},
    \label{eq:C02_eq}
\end{equation}
shows that $C_{02}=0$ requires $JC_{11}=-F_2C_{01}$: the direct drive
$F_2|0,1\rangle\!\to\!|0,2\rangle$ is exactly cancelled by hopping
$J|1,1\rangle\!\to\!|0,2\rangle$.  This cancellation fixes $C_{11}$
regardless of how $|1,1\rangle$ was reached.  Substituting $C_{11}$
into~\eqref{eq:A} gives $C_{20}$; inserting both into~\eqref{eq:B} and
imposing self-consistency of the one-photon ratio $r$ then yields
the \emph{main condition}:
\begin{equation}
    (\tilde{E}+U)(\tilde{E}+2iJ) = J^2.
    \label{eq:main_result}
\end{equation}
The factor $2iJ$ has two origins: the $\sqrt{2}J$ two-photon hopping matrix
elements (giving the factor of 2) and the $90^\circ$ drive phase $F_2=iF_1$
(giving the factor of $i$), which rotates the effective coupling into the
imaginary direction.  The simplification from the general quartic of
Ref.~\cite{shen_exact_2015} to a quadratic relies on $E_2=2\tilde{E}$, which
holds only for identical cavities.

\subsection{Solving for the Optimal Parameter Locus}

Writing $\tilde{E}=\Delta-i\gamma/2$ with $\Delta$ real and imposing
$\mathrm{Im}\bigl[E_U(\tilde{E}+2iJ)\bigr]=0$ (since the right-hand side
$J^2$ is real), one obtains
\begin{equation}
    U = \frac{2\Delta(\gamma-2J)}{4J-\gamma}.
    \label{eq:U_of_Delta}
\end{equation}
Substituting into the real part of~\eqref{eq:main_result} and solving for
$\Delta$ gives the closed-form locus (Eqs.~\ref{eq:Delta}--\ref{eq:U} of the main text):
\begin{align}
    \Delta_\mathrm{opt} &= \Bigl(\tfrac{\gamma}{2}-J\Bigr)\sqrt{\tfrac{4J}{\gamma}-1},
    \label{eq:Delta_locus}\\
    U_\mathrm{opt} &= \frac{4(J-\gamma/2)^2}{\sqrt{4J/\gamma-1}},
    \label{eq:U_locus}
\end{align}
valid for $J>J_\mathrm{min}=\gamma/4$.  Selected numerical values are
given in Table~\ref{tab:locus}.

\begin{table}[h]
\renewcommand{\arraystretch}{1.4}
\centering
\begin{tabular}{rrrr}
\toprule
$J/\gamma$ & $U_\mathrm{opt}/\gamma$ & $U_\mathrm{opt}/J$ & $\Delta_\mathrm{opt}/\gamma$ \\
\midrule
0.30 & 0.358 & 1.193 &  $+0.089$ \\
0.40 & 0.052 & 0.129 &  $+0.077$ \\
0.60 & 0.034 & 0.056 & $-0.118$ \\
0.70 & 0.119 & 0.170 & $-0.268$ \\
0.80 & 0.243 & 0.303 & $-0.445$ \\
1.00 & 0.577 & 0.577 & $-0.866$ \\
\bottomrule
\end{tabular}

\begin{minipage}{0.45\textwidth}
\caption{Parameter locus for the $\phi=90^\circ$ bilateral-drive UPB.
The $J=0.30\gamma$ row has $U/J>1$, which lies outside the spirit of
weak-nonlinearity UPB but still satisfies the interference condition exactly.}
\label{tab:locus}
\end{minipage}
\end{table}

% ============================================================
\section{General Drive Phase: Derivation and Phase Range}
\label{sec:general_phi}
% ============================================================

\subsection{General $C_{02}=0$ Condition for Arbitrary $\phi$}

For $F_2=F_1 e^{i\phi}$ with $p\equiv e^{i\phi}$, the one-photon
amplitudes are
\begin{equation}
    C_{10} = \frac{F_1(pJ-\tilde{E})}{\tilde{E}^2-J^2}, \qquad
    C_{01} = \frac{F_1(J-p\tilde{E})}{\tilde{E}^2-J^2},
    \label{eq:C10C01_general}
\end{equation}
with ratio $r=(J-p\tilde{E})/(pJ-\tilde{E})$.  Following the same
two-photon projection procedure as Appendix~\ref{sec:derivation} and using
$E_2=2\tilde{E}$, one derives
\begin{equation}
    J^2\bigl(2\tilde{E}+U(1+p^2)\bigr)
    = 2p\tilde{E}\,E_U(2J-p\tilde{E}),
    \label{eq:general_cond}
\end{equation}
solved explicitly for $U$ as
\begin{equation}
    U = \frac{-2\tilde{E}(J-p\tilde{E})^2}
             {J^2(1+p^2)-4pJ\tilde{E}+2p^2\tilde{E}^2}.
    \label{eq:U_phi_app}
\end{equation}
Setting $p=i$ makes $1+p^2=0$ and the denominator reduces to
$-2\tilde{E}(\tilde{E}+2iJ)$, recovering~\eqref{eq:main_result} exactly.

\subsection{Failure at $\phi=0$ and $\phi=\pi$}

At $p=1$ ($\phi=0$), Eq.~\eqref{eq:U_phi_app} gives $U=-\tilde{E}$, which
requires $\mathrm{Im}(U)=\gamma/2\neq0$ — no physical solution exists.
The same holds at $p=-1$ ($\phi=\pi$).  Destructive interference in the
two-photon sector demands a non-zero quadrature component; a purely
in-phase or anti-phase drive cannot produce $C_{02}=0$.

\subsection{Valid Phase Range and Minimum of $U_\mathrm{opt}$}

Physical solutions ($U$ real and positive) exist only in a finite range of
$\phi$ that is not symmetric about $90^\circ$.  For $J=0.4\gamma$ this range
is approximately $38^\circ$ to $151^\circ$.

The minimum of $U_\mathrm{opt}(\phi)$ at fixed $J$ occurs at
$\phi^*\approx92.2^\circ$ for $J=0.4\gamma$, only $0.7\%$ below the value
at $\phi=90^\circ$.  The phase $\phi^*$ approaches $90^\circ$ as $J\to(\gamma/2)^-$.
Table~\ref{tab:phi_tradeoff} lists $U_\mathrm{opt}/\gamma$ and
$\Delta_\mathrm{opt}/\gamma$ for representative phases at $J=0.4\gamma$.

\begin{table}[h]
\renewcommand{\arraystretch}{1.4}
\centering
\begin{tabular}{rrr}
\toprule
$\phi$ & $U_\mathrm{opt}/\gamma$ & $\Delta_\mathrm{opt}/\gamma$ \\
\midrule
$40^\circ$  & 1.835 & $+0.453$ \\
$60^\circ$  & 0.192 & $+0.317$ \\
$90^\circ$  & 0.052 & $+0.077$ \\
$120^\circ$ & 0.113 & $-0.112$ \\
$150^\circ$ & 0.663 & $-0.351$ \\
\bottomrule
\end{tabular}

\begin{minipage}{0.45\textwidth}
\caption{UPB operating points as a function of drive phase for
$J=0.4\gamma$.  Values from Eq.~\eqref{eq:U_phi_app} with $\mathrm{Im}(U)=0$.}
\label{tab:phi_tradeoff}
\end{minipage}
\end{table}

% ============================================================
\section{The Linear Dark State}
\label{sec:dark_state}
% ============================================================

For $J>\gamma/2$, Eq.~\eqref{eq:U_phi_app} admits a spurious solution with
$U=0$ at the phase $\sin\phi=\gamma/(2J)$.  At this point
$J=p\tilde{E}$, which causes $C_{01}=0$ exactly from~\eqref{eq:C10C01_general}:
the direct drive on site~2 is cancelled by the hopping amplitude, leaving
site~2 unpopulated at the \emph{single-photon} level.  Consequently
$C_{02}=0$ follows trivially — there is no photon population on site~2
to build on — and $g^{(2)}_{22}(0)=0/0$ is undefined rather than a
quantum antibunching signature.

% ============================================================
\section{Analytic $g^{(2)}_{22}(\tau)$ via the Quantum Regression Theorem}
\label{sec:g2_analytic}
% ============================================================

\subsection{Post-Detection State}

At the UPB operating point $C_{02}=0$, the state immediately after
detecting a photon from site~2 is, in the weak-drive limit:
\begin{equation}
    |\tilde{\Psi}(0)\rangle = C_{01}|0,0\rangle + C_{11}|1,0\rangle,
\end{equation}
which lies entirely in the zero- and one-photon sectors.  Hence
$g^{(2)}_{22}(0)=0$ exactly at the UPB point.

\subsection{Equations of Motion and Diagonalisation}

Under the CW bilateral drive, the one-photon amplitudes $C_{10}(\tau)$
and $C_{01}(\tau)$ obey the driven linear system~\eqref{eq:1ph_system}
(in time-dependent form with the vacuum amplitude fixed at $C_{01}$
following Ref.~\cite{shen_exact_2015}).  Defining sum and difference variables
$x=C_{10}+C_{01}$ and $y=C_{10}-C_{01}$, these decouple into two
first-order ODEs with decay rates and oscillation frequencies:
\begin{equation}
    u_{1,2} = i(\tilde{E}\pm J) = \tfrac{\gamma}{2}\mp i(\Delta\pm J).
    \label{eq:u12}
\end{equation}
Both modes decay since $\mathrm{Re}(u_{1,2})=\gamma/2>0$.
In the overdamped regime $J<\gamma/2$, the imaginary parts of $u_{1,2}$
have opposite signs and the two exponentials interfere, producing a transient with at most a single small overshoot of unity and no oscillations.

\subsection{Closed-Form Result}

With initial conditions $C_{10}(0)=C_{11}$, $C_{01}(0)=0$, the
solution is
\begin{equation}
\begin{split}
    C_{01}(\tau) = \frac{1}{2}\bigg[
    &\left(C_{11}-\frac{\lambda_1}{u_1}\right)e^{-u_1\tau} \\
    &-\left(C_{11}-\frac{\lambda_2}{u_2}\right)e^{-u_2\tau}
    +\frac{\lambda_1}{u_1}-\frac{\lambda_2}{u_2}\bigg],
    \label{eq:C01_final}
\end{split}
\end{equation}
where $\lambda_{1,2}=-i(F_1\pm F_2)C_{01}$.  The final result
$g^{(2)}_{22}(\tau)=|C_{01}(\tau)|^2/|C_{01}|^4$ gives $g^{(2)}_{22}(0)=0$ and $g^{(2)}_{22}(\infty)=1$ exactly,
as verified by substituting the steady-state relations
$(\tilde{E}\pm J)(C_{10}\pm C_{01})=-(F_1\pm F_2)$.

\subsection{Oscillation Frequencies at the UPB Locus}

Substituting $\Delta_\mathrm{opt}$ from~\eqref{eq:Delta_locus}, the
oscillation frequencies become:
\begin{equation}
    \omega_{1,2} = \Bigl(\tfrac{\gamma}{2}-J\Bigr)\sqrt{\tfrac{4J}{\gamma}-1}\pm J.
\end{equation}
For $J<\gamma/2$, the arguments of both exponentials in~\eqref{eq:C01_final}
have positive real parts, confirming overdamped, oscillation-free behaviour.

\subsection{Analytic $g^{(2)}_{11}(\tau)$}
 
For completeness, $g^{(2)}_{11}(\tau)$ is obtained analogously by applying
the quantum regression theorem after a site-1 detection.  The post-detection
state is $C_{10}|0,0\rangle+\sqrt{2}C_{20}|1,0\rangle+C_{11}|0,1\rangle$,
giving initial conditions $C_{10}(0)=\sqrt{2}C_{20}$, $C_{01}(0)=C_{11}$.
The result is $g^{(2)}_{11}(\tau)=|C_{10}(\tau)|^2/|C_{10}|^4$, where
\begin{equation}
\begin{split}
    C_{10}(\tau) = \frac{1}{2}\bigg[
    &\left(x_0-\frac{\lambda_1^{(11)}}{u_1}\right)e^{-u_1\tau} \\
    &+\left(y_0-\frac{\lambda_2^{(11)}}{u_2}\right)e^{-u_2\tau}
    +\frac{\lambda_1^{(11)}}{u_1}+\frac{\lambda_2^{(11)}}{u_2}\bigg],
\end{split}
\end{equation}
with $\lambda_{1,2}^{(11)}=-i(F_1\pm F_2)C_{10}$,
$x_0=\sqrt{2}C_{20}+C_{11}$, $y_0=\sqrt{2}C_{20}-C_{11}$.
At our $C_{02}=0$ operating point, $C_{20}\neq0$.  From
equation~\eqref{eq:A} with $C_{11}=-F_2C_{01}/J$ and $F_2=iF_1$,
$C_{01}=rC_{10}$:
\begin{equation}
    C_{20} = \frac{F_1 C_{10}(ir-1)}{\sqrt{2}\,(\tilde{E}+U)}.
\end{equation}
Substituting into $g^{(2)}_{11}(0)=2|C_{20}|^2/|C_{10}|^4$ using the
initial condition $C_{10}(0)=\sqrt{2}C_{20}$:
\begin{equation}
    g^{(2)}_{11}(0) = \frac{2|C_{20}|^2}{|C_{10}|^4}
    = \frac{|ir-1|^2\,F_1^2}{|\tilde{E}+U|^2\,|C_{10}|^2}.
\end{equation}
Numerically at the UPB optimum ($J=0.4\gamma$), this gives
$g^{(2)}_{11}(0)\approx0.98$, confirming that antibunching is selective
to site~2 while site~1 remains near-Poissonian.
% ============================================================
\section{Numerical Methods}
\label{sec:numerics}
% ============================================================

All numerical results were obtained using QuTiP~4 (Quantum Toolbox in
Python)~\cite{johansson_qutip_2013} in Python~3.  The Hilbert space was
truncated at $N_\mathrm{cut}=7$ photons per site (total dimension
$49$), verified to be sufficient by checking convergence at
$N_\mathrm{cut}=15$.

\textbf{Steady-state correlators.}  Equal-time correlators
$g^{(2)}_{jk}(0)$ and mean occupations $\langle n_j\rangle$ were
computed from the steady-state density matrix $\rho_\mathrm{ss}$
obtained via \texttt{qutip.steadystate} using the direct sparse LU
solver.  For the parameter landscape (Fig.~\ref{fig:landscape}) the grid
was evaluated in parallel across all CPU cores using Python's
\texttt{multiprocessing} module.  In the weak-drive limit
$F_1\ll\gamma$, the perturbative amplitude solver (Appendix~\ref{sec:derivation}) was used
instead for speed, with the full master equation reserved for
verification.

\textbf{Time-delayed correlators.}  $g^{(2)}_{jk}(\tau)$ was computed
via the quantum regression theorem (QRT) implemented directly with
\texttt{qutip.mesolve}.  For each correlator, the post-measurement state
$\hat{a}_k\,\rho_\mathrm{ss}\,\hat{a}_k^\dagger$ was normalised and
propagated forward in $\tau$ under the full Lindblad master equation;
the expectation value of $\hat{a}_j^\dagger\hat{a}_j$ on the propagated
state then yields $g^{(2)}_{jk}(\tau)$.

\textbf{Pulsed excitation.}  The time-dependent master equation with
a Gaussian drive envelope was integrated using the variable-step
Adams--Bashforth--Moulton solver in \texttt{qutip.mesolve}.
$g^{(2)}_{22}(\tau)$ was then computed via the QRT with $\tau=0$
referenced to the pulse peak.

\begin{figure}[h]
\centering
\includegraphics[width=\columnwidth]{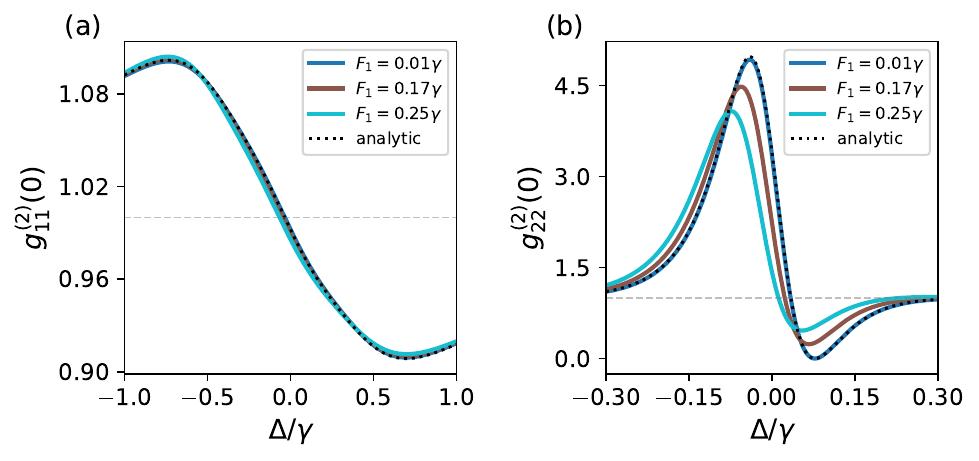}
\caption{Equal-time correlators $g^{(2)}_{11}(0)$ (left) and
$g^{(2)}_{22}(0)$ (right) vs detuning $\Delta/\gamma$ at
$J=0.4\gamma$, $U=0.052\gamma$, $\phi=90^\circ$, for three drive
amplitudes $F_1/\gamma\in\{0.01,\,0.17,\,0.25\}$ (solid coloured
curves, full master equation).  The black dotted curve is the analytic
no-jump result, which is independent of $F_1$ in the weak-drive limit.}
\label{fig:landscape_cuts}
\end{figure}

\textbf{Validity of the no-jump approximation.}
Figure~\ref{fig:landscape_cuts} compares the full master-equation result
for $g^{(2)}_{11}(0)$ and $g^{(2)}_{22}(0)$ as a function of detuning
$\Delta$ at three drive amplitudes ($F_1/\gamma = 0.01$, $0.17$, $0.25$)
with the analytic no-jump prediction (black dotted curve), which is
independent of $F_1$ in the weak-drive limit.  At $F_1=0.01\gamma$ the
master-equation result is indistinguishable from the analytic curve.
As $F_1$ increases, the minimum of $g^{(2)}_{22}(0)$ rises and redshifts, reflecting corrections of order $(F_1/\gamma)^2$ beyond the
no-jump approximation.
 
% ============================================================
\section{Peak Overshoot of $g^{(2)}_{22}(\tau)$ Across All $J$}
\label{sec:osc_amplitude}
% ============================================================
\begin{figure}[t]
\centering
\includegraphics[width=\columnwidth]{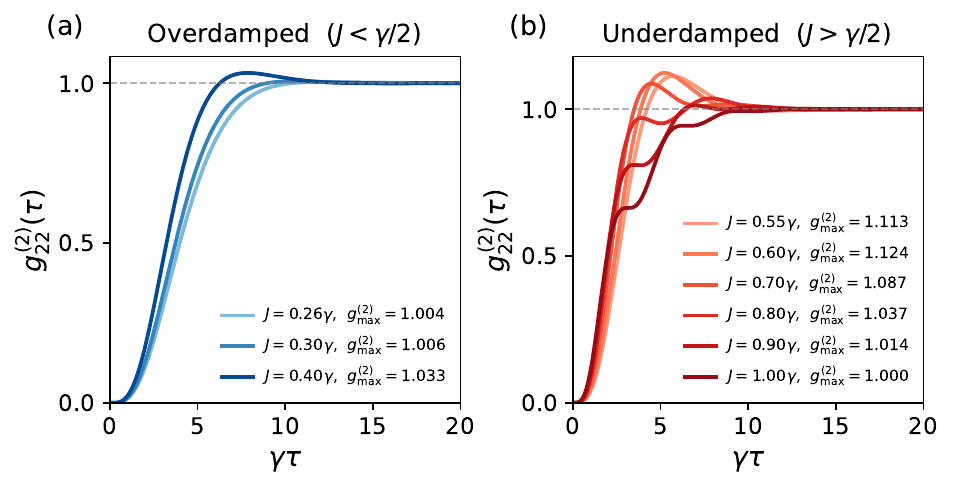}
\caption{Analytic $g^{(2)}_{22}(\tau)$ at the UPB optimum ($\phi=90^\circ$,
$F_1=0.05\gamma$) for various $J/\gamma$.
(a) \textit{Blue panels} ($J<\gamma/2$): smooth overdamped rise, overshooting
unity by at most $8\%$. (b) \textit{Red panels} ($J>\gamma/2$): visible
oscillations at frequencies $\omega_{1,2}=\Delta_\mathrm{opt}\pm J$,
with peak overshoot decreasing monotonically to $0.04\%$ at $J=\gamma$.}
\label{fig:g22_supplementary}
\end{figure}

Using the analytic formula~\eqref{eq:C01_final}, Table~\ref{tab:osc_freqs}
gives the oscillation frequencies and numerically computed peak value
$g^{(2)}_\mathrm{max}\equiv\max_{\tau>0} g^{(2)}_{22}(\tau)$ across the full
valid range $J>\gamma/4$.

\begin{table}[h]
\renewcommand{\arraystretch}{1.4}
\centering
\begin{tabular}{@{}rrrrrrr@{}}
\toprule
$J/\gamma$ & $\Delta_\mathrm{opt}/\gamma$ & $U_\mathrm{opt}/\gamma$
           & $\omega_1/\gamma$ & $\omega_2/\gamma$
           & $T_2\gamma$ & $g^{(2)}_\mathrm{max}$ \\
\midrule
0.26 & $+0.048$ & 1.152 & $+0.308$ & $-0.212$ & 29.7 & 1.004 \\
0.30 & $+0.089$ & 0.358 & $+0.389$ & $-0.211$ & 29.8 & 1.006 \\
0.40 & $+0.077$ & 0.052 & $+0.478$ & $-0.323$ & 19.5 & 1.033 \\
\midrule
0.55 & $-0.055$ & 0.009 & $+0.495$ & $-0.605$ & 10.4 & 1.113 \\
0.60 & $-0.118$ & 0.034 & $+0.482$ & $-0.718$ &  8.7 & 1.124 \\
0.70 & $-0.268$ & 0.119 & $+0.432$ & $-0.968$ &  6.5 & 1.087 \\
0.80 & $-0.445$ & 0.243 & $+0.355$ & $-1.245$ &  5.0 & 1.037 \\
0.90 & $-0.645$ & 0.397 & $+0.255$ & $-1.545$ &  4.1 & 1.014 \\
1.00 & $-0.866$ & 0.577 & $+0.134$ & $-1.866$ &  3.4 & 1.000 \\
\bottomrule
\end{tabular}
\begin{minipage}{0.45\textwidth}
\caption{Oscillation frequencies and peak overshoot of $g^{(2)}_{22}(\tau)$
at the $\phi=90^\circ$ UPB optimum.  $T_{1,2}=2\pi/|\omega_{1,2}|$ are the
oscillation periods.  The row at $J=\gamma/2$ is the transition between
overdamped and underdamped regimes; $U_\mathrm{opt}=0$ at this point
corresponds to the linear dark state and is excluded in practice.}
\label{tab:osc_freqs}
\end{minipage}
\end{table}

The peak overshoot $g^{(2)}_\mathrm{max}-1$ is non-monotonic.  In the
overdamped regime ($J<\gamma/2$) it grows from $\sim0.4\%$ near threshold
to $\sim8\%$ approaching $J=\gamma/2$.  In the underdamped regime
($J>\gamma/2$) it decreases monotonically, reaching $0.04\%$ at
$J=\gamma$ despite oscillation periods $T_2\sim3/\gamma$.  The shrinking
overshoot reflects decreasing transient amplitudes $(C_{11}-\lambda/u)$
in~\eqref{eq:C01_final} relative to the steady-state $|C_{01}|^2$ as
$J$ grows.

Figure~\ref{fig:g22_supplementary} shows the analytic $g^{(2)}_{22}(\tau)$
across the full valid range, illustrating the crossover from overdamped
(smooth) to underdamped (oscillatory) behaviour.

% ============================================================
\section{Disorder Robustness}
\label{sec:disorder}
% ============================================================

We assess the sensitivity of $g^{(2)}_{22}(0)$ to site-to-site parameter
mismatch using the perturbative amplitude solver extended to asymmetric
cavities.  Three disorder axes are explored independently at the analytic
optimum ($J=0.4\gamma$, $U=0.05164\gamma$, $\Delta=0.07746\gamma$,
$\phi=90^\circ$, $F_1=0.05\gamma$): detuning mismatch
$\delta\Delta=\Delta_1-\Delta_2$, loss-rate mismatch
$\delta\gamma=\gamma_1-\gamma_2$, and nonlinearity mismatch
$\delta U=U_1-U_2$. We define the tolerance as the mismatch at which 
$g^{(2)}_{22}(0)$ exceeds $0.1$.

% \begin{center}
% \renewcommand{\arraystretch}{1.4}
% \begin{tabular}{lr}
% \toprule
% Disorder axis & Tolerance \\
% \midrule
% Detuning $\delta\Delta$       & $\pm0.033\,\gamma$ \\
% Loss rate $\delta\gamma$      & $\pm0.060\,\gamma$ \\
% Nonlinearity $\delta U$       & $\pm0.033\,\gamma$ \\
% \bottomrule
% \end{tabular}
% \end{center}

\begin{table}[h]
\renewcommand{\arraystretch}{1.4}
\centering
\begin{tabular}{lr}
\toprule
Disorder axis & Tolerance \\
\midrule
Detuning $\delta\Delta$       & $\pm0.033\,\gamma$ \\
Loss rate $\delta\gamma$      & $\pm0.060\,\gamma$ \\
Nonlinearity $\delta U$       & $\pm0.033\,\gamma$ \\
\bottomrule
\end{tabular}
\caption{Disorder tolerances at the UPB optimum without compensation.}
\label{tab:tolerances}
\end{table}

The tightest constraint is the detuning tolerance $\pm0.033\,\gamma$ (Table~\ref{tab:tolerances}).
For a GaAs photonic molecule with $Q\simeq10^5$ at $\lambda=850\,\text{nm}$
($\gamma\approx25\,\text{GHz}$), this corresponds to a frequency mismatch
of $\sim0.8\,\text{GHz}$ ($\sim3\,\mu\text{eV}$).  The loss-rate
mismatch tolerance is roughly twice as large.

% \begin{figure*}
% \centering
% \includegraphics[width=0.7\textwidth]{disorder_robustness.pdf}
% \caption{Disorder robustness at the UPB optimum
% ($J=0.4\gamma$, $U=0.05164\gamma$, $\Delta=0.07746\gamma$, $\phi=90^\circ$).
% (a)~$g^{(2)}_{22}(0)$ vs detuning mismatch $\delta\Delta$.
% (b)~$g^{(2)}_{22}(0)$ vs loss-rate mismatch $\delta\gamma$.
% (c)~$g^{(2)}_{22}(0)$ vs nonlinearity mismatch $\delta U$.
% Dashed red lines mark the $g^{(2)}_{22}=0.1$ threshold.
% (d)~2-D map over $(\delta\Delta,\delta\gamma)$; contour at $g^{(2)}_{22}=0.1$.
% (e)~Tolerance half-widths for each disorder axis.}
% \label{fig:disorder}
% \end{figure*}

\subsection{Compensation of disorder by drive re-tuning}

\begin{figure}
\centering
\includegraphics[width=\columnwidth]{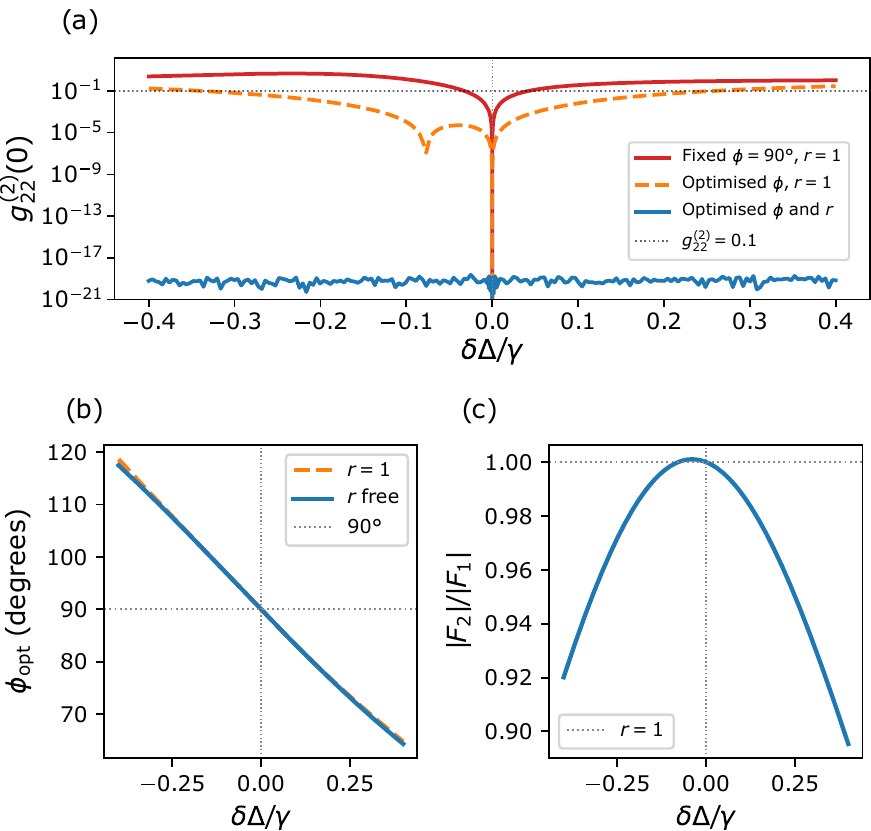}
\caption{Drive-parameter compensation of detuning disorder
($J=0.4\gamma$, $U=0.05164\gamma$, $\Delta=0.07746\gamma$).
(a)~$g^{(2)}_{22}(0)$ vs $\delta\Delta$ for three cases: fixed
$\phi=90^\circ$ and $r=1$ (red), optimised $\phi$ with $r=1$ (orange dashed),
and optimised $\phi$ and $r$ (blue).  Dotted line: $g^{(2)}_{22}=0.1$
threshold.
(b)~Optimal $\phi$ as a function of $\delta\Delta$ for both compensated cases.
(c)~Optimal amplitude ratio $r=|F_2|/|F_1|$ for full compensation.}
\label{fig:compensation}
\end{figure}

The tolerances above assume fixed drive parameters.  Since the $C_{02}=0$
condition is one complex equation in three real system parameters, adding
site-to-site disorder introduces new unknowns that can in principle be
absorbed by re-tuning the drive.  We investigate two strategies: (i)
adjusting the drive phase $\phi$ alone while keeping $|F_2|=|F_1|$, and
(ii) adjusting both $\phi$ and the amplitude ratio $r=|F_2|/|F_1|$.

For each value of detuning mismatch $\delta\Delta$, we minimise
$g^{(2)}_{22}(0)$ numerically over the drive parameters.
Figure~\ref{fig:compensation} shows the result.  Optimising $\phi$ alone
extends the tolerance from $\pm0.033\,\gamma$ to $\pm0.26\,\gamma$ — nearly
an order of magnitude — because $\phi$ directly controls the relative phase
between the direct and indirect two-photon pathways.  Optimising both $\phi$
and $r$ achieves $g^{(2)}_{22}(0)\lesssim10^{-18}$ across the entire scanned
range $|\delta\Delta|\leq0.4\,\gamma$, demonstrating that the UPB condition
can be exactly restored for arbitrary detuning disorder by drive re-tuning
alone, with no need for post-fabrication cavity trimming.

Panel~(b) shows that the compensating $\phi$ tracks the disorder nearly
linearly, deviating from $90^\circ$ at a rate of approximately $15^\circ$ per
$0.1\,\gamma$ of mismatch.  When $r$ is also free, it decreases smoothly
from 1 as the mismatch grows (panel~(c)), indicating that the two
pathways require unequal drive amplitudes to maintain destructive
interference in the asymmetric dimer.

\newpage
\clearpage

\bibliography{refs}

@article{michler_quantum_2000,
    title = {A {Quantum} {Dot} {Single}-{Photon} {Turnstile} {Device}},
    volume = {290},
    issn = {0036-8075, 1095-9203},
    url = {https://www.science.org/doi/10.1126/science.290.5500.2282},
    doi = {10.1126/science.290.5500.2282},
    language = {en},
    number = {5500},
    urldate = {2026-04-06},
    journal = {Science},
    author = {Michler, P. and Kiraz, A. and Becher, C. and Schoenfeld, W. V. and Petroff, P. M. and Zhang, Lidong and Hu, E. and Imamoglu, A.},
    month = dec,
    year = {2000},
    pages = {2282--2285},
}

@article{senellart_high-performance_2017,
    title = {High-performance semiconductor quantum-dot single-photon sources},
    volume = {12},
    issn = {1748-3387, 1748-3395},
    url = {https://www.nature.com/articles/nnano.2017.218},
    doi = {10.1038/nnano.2017.218},
    language = {en},
    number = {11},
    urldate = {2026-04-06},
    journal = {Nature Nanotechnology},
    author = {Senellart, Pascale and Solomon, Glenn and White, Andrew},
    month = nov,
    year = {2017},
    pages = {1026--1039},
}

@article{tian_quantum_1992,
    title = {Quantum trajectory simulations of two-state behavior in an optical cavity containing one atom},
    volume = {46},
    copyright = {http://link.aps.org/licenses/aps-default-license},
    issn = {1050-2947, 1094-1622},
    url = {https://link.aps.org/doi/10.1103/PhysRevA.46.R6801},
    doi = {10.1103/PhysRevA.46.R6801},
    language = {en},
    number = {11},
    urldate = {2026-04-06},
    journal = {Physical Review A},
    author = {Tian, L. and Carmichael, H. J.},
    month = dec,
    year = {1992},
    pages = {R6801--R6804},
}

@article{imamoglu_strongly_1997,
    title = {Strongly {Interacting} {Photons} in a {Nonlinear} {Cavity}},
    volume = {79},
    url = {https://link.aps.org/doi/10.1103/PhysRevLett.79.1467},
    doi = {10.1103/PhysRevLett.79.1467},
    number = {8},
    urldate = {2018-03-30},
    journal = {Physical Review Letters},
    author = {Imamoḡlu, A. and Schmidt, H. and Woods, G. and Deutsch, M.},
    month = aug,
    year = {1997},
    pages = {1467--1470},
}

@article{birnbaum_photon_2005,
    title = {Photon blockade in an optical cavity with one trapped atom},
    volume = {436},
    copyright = {2005 Springer Nature Limited},
    issn = {1476-4687},
    url = {https://www.nature.com/articles/nature03804},
    doi = {10.1038/nature03804},
    language = {en},
    number = {7047},
    urldate = {2025-06-23},
    journal = {Nature},
    publisher = {Nature Publishing Group},
    author = {Birnbaum, K. M. and Boca, A. and Miller, R. and Boozer, A. D. and Northup, T. E. and Kimble, H. J.},
    month = jul,
    year = {2005},
    pages = {87--90},
}

@article{faraon_coherent_2008,
    title = {Coherent generation of non-classical light on a chip via photon-induced tunnelling and blockade},
    volume = {4},
    copyright = {2008 Springer Nature Limited},
    issn = {1745-2481},
    url = {https://www.nature.com/articles/nphys1078},
    doi = {10.1038/nphys1078},
    language = {en},
    number = {11},
    urldate = {2025-06-23},
    journal = {Nature Physics},
    publisher = {Nature Publishing Group},
    author = {Faraon, Andrei and Fushman, Ilya and Englund, Dirk and Stoltz, Nick and Petroff, Pierre and Vučković, Jelena},
    month = nov,
    year = {2008},
    pages = {859--863},
}

@article{lang_observation_2011,
    title = {Observation of {Resonant} {Photon} {Blockade} at {Microwave} {Frequencies} {Using} {Correlation} {Function} {Measurements}},
    volume = {106},
    copyright = {http://link.aps.org/licenses/aps-default-license},
    issn = {0031-9007, 1079-7114},
    url = {https://link.aps.org/doi/10.1103/PhysRevLett.106.243601},
    doi = {10.1103/PhysRevLett.106.243601},
    language = {en},
    number = {24},
    urldate = {2026-04-06},
    journal = {Physical Review Letters},
    author = {Lang, C. and Bozyigit, D. and Eichler, C. and Steffen, L. and Fink, J. M. and Abdumalikov, A. A. and Baur, M. and Filipp, S. and Da Silva, M. P. and Blais, A. and Wallraff, A.},
    month = jun,
    year = {2011},
    pages = {243601},
}

@article{bamba_origin_2011,
    title = {Origin of strong photon antibunching in weakly nonlinear photonic molecules},
    volume = {83},
    url = {https://link.aps.org/doi/10.1103/PhysRevA.83.021802},
    doi = {10.1103/PhysRevA.83.021802},
    number = {2},
    urldate = {2019-07-31},
    journal = {Physical Review A},
    author = {Bamba, Motoaki and Imamoğlu, Atac and Carusotto, Iacopo and Ciuti, Cristiano},
    month = feb,
    year = {2011},
    pages = {021802},
}

@article{orfanakis_rydberg_2022,
    title = {Rydberg exciton–polaritons in a {Cu2O} microcavity},
    volume = {21},
    copyright = {2022 The Author(s), under exclusive licence to Springer Nature Limited},
    issn = {1476-4660},
    url = {http://www.nature.com/articles/s41563-022-01230-4},
    doi = {10.1038/s41563-022-01230-4},
    language = {en},
    number = {7},
    urldate = {2022-06-30},
    journal = {Nature Materials},
    publisher = {Nature Publishing Group},
    author = {Orfanakis, Konstantinos and Rajendran, Sai Kiran and Walther, Valentin and Volz, Thomas and Pohl, Thomas and Ohadi, Hamid},
    month = jul,
    year = {2022},
    note = {Number: 7},
    pages = {767--772},
}

@article{liew_single_2010,
    title = {Single {Photons} from {Coupled} {Quantum} {Modes}},
    volume = {104},
    issn = {0031-9007, 1079-7114},
    url = {https://link.aps.org/doi/10.1103/PhysRevLett.104.183601},
    doi = {10.1103/PhysRevLett.104.183601},
    language = {en},
    number = {18},
    urldate = {2019-07-30},
    journal = {Physical Review Letters},
    author = {Liew, T. C. H. and Savona, V.},
    month = may,
    year = {2010},
    pages = {183601},
}

@article{ferretti_photon_2010,
    title = {Photon correlations in a two-site nonlinear cavity system under coherent drive and dissipation},
    volume = {82},
    copyright = {http://link.aps.org/licenses/aps-default-license},
    issn = {1050-2947, 1094-1622},
    url = {https://link.aps.org/doi/10.1103/PhysRevA.82.013841},
    doi = {10.1103/PhysRevA.82.013841},
    language = {en},
    number = {1},
    urldate = {2026-04-06},
    journal = {Physical Review A},
    author = {Ferretti, Sara and Andreani, Lucio Claudio and Türeci, Hakan E. and Gerace, Dario},
    month = jul,
    year = {2010},
    pages = {013841},
}

@article{flayac_all-silicon_2015,
    title = {An all-silicon single-photon source by unconventional photon blockade},
    volume = {5},
    issn = {2045-2322},
    url = {https://www.nature.com/articles/srep11223},
    doi = {10.1038/srep11223},
    language = {en},
    number = {1},
    urldate = {2026-04-06},
    journal = {Scientific Reports},
    author = {Flayac, Hugo and Gerace, Dario and Savona, Vincenzo},
    month = jun,
    year = {2015},
    pages = {11223},
}

@article{shen_exact_2015,
    title = {Exact optimal control of photon blockade with weakly nonlinear coupled cavities},
    volume = {23},
    copyright = {© 2015 Optical Society of America},
    issn = {1094-4087},
    url = {https://opg.optica.org/oe/abstract.cfm?uri=oe-23-25-32835},
    doi = {10.1364/OE.23.032835},
    language = {EN},
    number = {25},
    urldate = {2026-03-27},
    journal = {Optics Express},
    publisher = {Optica Publishing Group},
    author = {Shen, H. Z. and Zhou, Y. H. and Liu, H. D. and Wang, G. C. and Yi, X. X.},
    month = dec,
    year = {2015},
    pages = {32835--32858},
}

@misc{wang_long-lived_2025,
    title = {Long-{Lived} {Photon} {Blockade} with {Weak} {Optical} {Nonlinearity}},
    url = {http://arxiv.org/abs/2502.09930},
    doi = {10.48550/arXiv.2502.09930},
    urldate = {2026-04-06},
    publisher = {arXiv},
    author = {Wang, You and Zheng, Xu and Liew, Timothy C. H. and Chong, Y. D.},
    month = jul,
    year = {2025},
    note = {arXiv:2502.09930 [quant-ph]},
}

@article{kyriienko_nonlinear_2020,
    title = {Nonlinear {Quantum} {Optics} with {Trion} {Polaritons} in {2D} {Monolayers}: {Conventional} and {Unconventional} {Photon} {Blockade}},
    volume = {125},
    issn = {0031-9007, 1079-7114},
    shorttitle = {Nonlinear {Quantum} {Optics} with {Trion} {Polaritons} in {2D} {Monolayers}},
    url = {https://link.aps.org/doi/10.1103/PhysRevLett.125.197402},
    doi = {10.1103/PhysRevLett.125.197402},
    language = {en},
    number = {19},
    urldate = {2026-04-06},
    journal = {Physical Review Letters},
    author = {Kyriienko, O. and Krizhanovskii, D. N. and Shelykh, I. A.},
    month = nov,
    year = {2020},
    pages = {197402},
}

@article{delteil_towards_2019,
    title = {Towards polariton blockade of confined exciton–polaritons},
    volume = {18},
    copyright = {2019 The Author(s), under exclusive licence to Springer Nature Limited},
    issn = {1476-4660},
    doi = {10.1038/s41563-019-0282-y},
    language = {En},
    number = {3},
    urldate = {2019-05-07},
    journal = {Nature Materials},
    publisher = {Nature Publishing Group},
    author = {Delteil, Aymeric and Fink, Thomas and Schade, Anne and Höfling, Sven and Schneider, Christian and İmamoğlu, Ataç},
    month = mar,
    year = {2019},
    note = {tex.ids= delteilPolaritonBlockadeConfined2019
number: 3},
    pages = {219},
}

@article{munoz-matutano_emergence_2019,
    title = {Emergence of quantum correlations from interacting fibre-cavity polaritons},
    volume = {18},
    copyright = {2019 Crown},
    issn = {1476-4660},
    url = {https://www.nature.com/articles/s41563-019-0281-z},
    doi = {10.1038/s41563-019-0281-z},
    language = {En},
    number = {3},
    urldate = {2019-05-07},
    journal = {Nature Materials},
    author = {Muñoz-Matutano, Guillermo and Wood, Andrew and Johnsson, Mattias and Vidal, Xavier and Baragiola, Ben Q. and Reinhard, Andreas and Lemaître, Aristide and Bloch, Jacqueline and Amo, Alberto and Nogues, Gilles and Besga, Benjamin and Richard, Maxime and Volz, Thomas},
    month = mar,
    year = {2019},
    pages = {213},
}

@article{zhang_van_2021,
    title = {Van der {Waals} heterostructure polaritons with moiré-induced nonlinearity},
    volume = {591},
    issn = {0028-0836, 1476-4687},
    url = {https://www.nature.com/articles/s41586-021-03228-5},
    doi = {10.1038/s41586-021-03228-5},
    language = {en},
    number = {7848},
    urldate = {2026-04-06},
    journal = {Nature},
    author = {Zhang, Long and Wu, Fengcheng and Hou, Shaocong and Zhang, Zhe and Chou, Yu-Hsun and Watanabe, Kenji and Taniguchi, Takashi and Forrest, Stephen R. and Deng, Hui},
    month = mar,
    year = {2021},
    pages = {61--65},
}

@article{flayac_single_2016,
    title = {Single photons from dissipation in coupled cavities},
    volume = {94},
    copyright = {http://link.aps.org/licenses/aps-default-license},
    issn = {2469-9926, 2469-9934},
    url = {https://link.aps.org/doi/10.1103/PhysRevA.94.013815},
    doi = {10.1103/PhysRevA.94.013815},
    language = {en},
    number = {1},
    urldate = {2026-04-06},
    journal = {Physical Review A},
    author = {Flayac, H. and Savona, V.},
    month = jul,
    year = {2016},
    pages = {013815},
}

@article{galbiati_polariton_2012,
    title = {Polariton {Condensation} in {Photonic} {Molecules}},
    volume = {108},
    url = {https://link.aps.org/doi/10.1103/PhysRevLett.108.126403},
    doi = {10.1103/PhysRevLett.108.126403},
    number = {12},
    journal = {Physical Review Letters},
    author = {Galbiati, Marta and Ferrier, Lydie and Solnyshkov, Dmitry D. and Tanese, Dimitrii and Wertz, Esther and Amo, Alberto and Abbarchi, Marco and Senellart, Pascale and Sagnes, Isabelle and Lemaître, Aristide and Galopin, Elisabeth and Malpuech, Guillaume and Bloch, Jacqueline},
    month = mar,
    year = {2012},
    pages = {126403},
}

@article{abbarchi_macroscopic_2013,
    title = {Macroscopic quantum self-trapping and {Josephson} oscillations of exciton polaritons},
    volume = {9},
    copyright = {© 2013 Nature Publishing Group},
    issn = {1745-2473},
    url = {http://www.nature.com/nphys/journal/v9/n5/full/nphys2609.html},
    doi = {10.1038/nphys2609},
    language = {en},
    number = {5},
    urldate = {2014-07-28},
    journal = {Nature Physics},
    author = {Abbarchi, M. and Amo, A. and Sala, V. G. and Solnyshkov, D. D. and Flayac, H. and Ferrier, L. and Sagnes, I. and Galopin, E. and Lemaître, A. and Malpuech, G. and Bloch, J.},
    month = may,
    year = {2013},
    pages = {275--279},
}

@article{reitzenstein_polarization-dependent_2010,
    title = {Polarization-dependent strong coupling in elliptical high-\${Q}\$ micropillar cavities},
    volume = {82},
    url = {https://link.aps.org/doi/10.1103/PhysRevB.82.235313},
    doi = {10.1103/PhysRevB.82.235313},
    number = {23},
    urldate = {2026-04-09},
    journal = {Physical Review B},
    publisher = {American Physical Society},
    author = {Reitzenstein, S. and Böckler, C. and Löffler, A. and Höfling, S. and Worschech, L. and Forchel, A. and Yao, P. and Hughes, S.},
    month = dec,
    year = {2010},
    pages = {235313},
}

@article{vladimirova_polariton-polariton_2010,
    title = {Polariton-polariton interaction constants in microcavities},
    volume = {82},
    url = {http://link.aps.org/doi/10.1103/PhysRevB.82.075301},
    doi = {10.1103/PhysRevB.82.075301},
    number = {7},
    urldate = {2012-09-25},
    journal = {Physical Review B},
    author = {Vladimirova, M. and Cronenberger, S. and Scalbert, D. and Kavokin, K. V. and Miard, A. and Lemaître, A. and Bloch, J. and Solnyshkov, D. and Malpuech, G. and Kavokin, A. V.},
    month = aug,
    year = {2010},
    pages = {075301},
}

@article{flayac_unconventional_2017,
    title = {Unconventional photon blockade},
    volume = {96},
    url = {https://link.aps.org/doi/10.1103/PhysRevA.96.053810},
    doi = {10.1103/PhysRevA.96.053810},
    number = {5},
    urldate = {2026-03-27},
    journal = {Physical Review A},
    publisher = {American Physical Society},
    author = {Flayac, H. and Savona, V.},
    month = nov,
    year = {2017},
    pages = {053810},
}

@article{couteau_applications_2023,
    title = {Applications of single photons to quantum communication and computing},
    volume = {5},
    copyright = {2023 Springer Nature Limited},
    issn = {2522-5820},
    url = {https://www.nature.com/articles/s42254-023-00583-2},
    doi = {10.1038/s42254-023-00583-2},
    language = {en},
    number = {6},
    urldate = {2026-04-09},
    journal = {Nature Reviews Physics},
    publisher = {Nature Publishing Group},
    author = {Couteau, Christophe and Barz, Stefanie and Durt, Thomas and Gerrits, Thomas and Huwer, Jan and Prevedel, Robert and Rarity, John and Shields, Andrew and Weihs, Gregor},
    month = jun,
    year = {2023},
    pages = {326--338},
}

@article{johansson_qutip_2013,
    title = {{QuTiP} 2: {A} {Python} framework for the dynamics of open quantum systems},
    volume = {184},
    issn = {0010-4655},
    shorttitle = {{QuTiP} 2},
    url = {https://www.sciencedirect.com/science/article/pii/S0010465512003955},
    doi = {10.1016/j.cpc.2012.11.019},
    number = {4},
    urldate = {2026-04-09},
    journal = {Computer Physics Communications},
    author = {Johansson, J. R. and Nation, P. D. and Nori, Franco},
    month = apr,
    year = {2013},
    pages = {1234--1240},
}

@article{chong_coherent_2010,
    title = {Coherent {Perfect} {Absorbers}: {Time}-{Reversed} {Lasers}},
    volume = {105},
    shorttitle = {Coherent {Perfect} {Absorbers}},
    url = {https://link.aps.org/doi/10.1103/PhysRevLett.105.053901},
    doi = {10.1103/PhysRevLett.105.053901},
    number = {5},
    urldate = {2026-04-16},
    journal = {Physical Review Letters},
    publisher = {American Physical Society},
    author = {Chong, Y. D. and Ge, Li and Cao, Hui and Stone, A. D.},
    month = jul,
    year = {2010},
    pages = {053901},
}

@article{zanotto_perfect_2014,
    title = {Perfect energy-feeding into strongly coupled systems and interferometric control of polariton absorption},
    volume = {10},
    copyright = {2014 Springer Nature Limited},
    issn = {1745-2481},
    url = {https://www.nature.com/articles/nphys3106},
    doi = {10.1038/nphys3106},
    language = {en},
    number = {11},
    urldate = {2026-04-16},
    journal = {Nature Physics},
    publisher = {Nature Publishing Group},
    author = {Zanotto, Simone and Mezzapesa, Francesco P. and Bianco, Federica and Biasiol, Giorgio and Baldacci, Lorenzo and Vitiello, Miriam Serena and Sorba, Lucia and Colombelli, Raffaele and Tredicucci, Alessandro},
    month = nov,
    year = {2014},
    pages = {830--834},
}

@article{wang_unconventional_2017,
    title = {Unconventional photon blockade in weakly nonlinear photonic molecules with bilateral drive},
    volume = {64},
    issn = {0950-0340},
    url = {https://doi.org/10.1080/09500340.2016.1249978},
    doi = {10.1080/09500340.2016.1249978},
    number = {6},
    urldate = {2026-06-17},
    journal = {Journal of Modern Optics},
    publisher = {Taylor \& Francis},
    author = {Wang, Gangcheng and Shen, H. Z. and Sun, Chunfang and Wu, Chunfeng and Chen, Jing-Ling and Xue, Kang},
    month = mar,
    year = {2017},
    note = {\_eprint: https://doi.org/10.1080/09500340.2016.1249978},
    pages = {583--590},
}

@article{shen_unconventional_2018,
    title = {Unconventional photon blockade from bimodal driving and dissipations in coupled semiconductor microcavities},
    volume = {51},
    issn = {0953-4075},
    url = {https://doi.org/10.1088/1361-6455/aa9c90},
    doi = {10.1088/1361-6455/aa9c90},
    language = {en},
    number = {3},
    urldate = {2026-06-17},
    journal = {Journal of Physics B: Atomic, Molecular and Optical Physics},
    publisher = {IOP Publishing},
    author = {Shen, H Z and Xu, Shuang and Zhou, Y H and Wang, Gangcheng and Yi, X X},
    month = jan,
    year = {2018},
    pages = {035503},
}

@article{zhou_universal_2025,
    title = {Universal {Photon} {Blockade}},
    volume = {134},
    url = {https://link.aps.org/doi/10.1103/PhysRevLett.134.183601},
    doi = {10.1103/PhysRevLett.134.183601},
    number = {18},
    urldate = {2026-06-17},
    journal = {Physical Review Letters},
    publisher = {American Physical Society},
    author = {Zhou, Yan-Hui and Liu, Tong and Su, Qi-Ping and Zhang, Xing-Yuan and Wu, Qi-Cheng and Chen, Dong-Xu and Shi, Zhi-Cheng and Shen, H. Z. and Yang, Chui-Ping},
    month = may,
    year = {2025},
    pages = {183601},
}

@article{flayac_input-output_2013,
    title = {Input-output theory of the unconventional photon blockade},
    volume = {88},
    url = {https://link.aps.org/doi/10.1103/PhysRevA.88.033836},
    doi = {10.1103/PhysRevA.88.033836},
    number = {3},
    urldate = {2026-06-18},
    journal = {Physical Review A},
    publisher = {American Physical Society},
    author = {Flayac, H. and Savona, V.},
    month = sep,
    year = {2013},
    pages = {033836},
}

@article{peng_chiral_2016,
    title = {Chiral modes and directional lasing at exceptional points},
    volume = {113},
    url = {https://www.pnas.org/doi/10.1073/pnas.1603318113},
    doi = {10.1073/pnas.1603318113},
    number = {25},
    urldate = {2026-06-17},
    journal = {Proceedings of the National Academy of Sciences},
    publisher = {Proceedings of the National Academy of Sciences},
    author = {Peng, Bo and Özdemir, Sahin Kaya and Liertzer, Matthias and Chen, Weijian and Kramer, Johannes and Yılmaz, Huzeyfe and Wiersig, Jan and Rotter, Stefan and Yang, Lan},
    month = jun,
    year = {2016},
    pages = {6845--6850},
}

\end{document}